\documentclass[12pt]{article}
\usepackage{amsmath}
\usepackage{graphicx,psfrag,epsf}
\usepackage{enumerate}
\usepackage{natbib}
\usepackage{url} 

\newcommand{\blind}{0}

\begin{document}

\def\spacingset#1{\renewcommand{\baselinestretch}%
{#1}\small\normalsize} \spacingset{1}


\if0\blind
{
  \title{\bf Parametric Sequential Causal Inference in Point Parametrization}
  \author{Li Yin \hspace{.2cm}\\
    Department of Medical Epidemiology and Biostatistics, Karolinska Institute\\
    and \\
    Xiaoqin Wang \\
    Department of Electronics, Mathematics and
Natural Sciences,  \\ University of G\"{a}vle}
  \maketitle
} \fi

\if1\blind
{
  \bigskip
  \bigskip
  \bigskip
  \begin{center}
    {\LARGE\bf Title}
\end{center}
  \medskip
} \fi

\bigskip
\begin{abstract}
Suppose that a sequence of treatments  are assigned to influence an outcome of interest that occurs after the last treatment. Between treatments there exist   time-dependent covariates  that may be posttreatment variables of the earlier treatments and confounders of the subsequent  treatments.
In this article, we develop a parametric approach to inference of the
causal effect of the treatment sequence on the outcome called the sequential causal effect.
We construct a
point parametrization for the conditional distribution of  an outcome given  all     treatments   and     time-dependent covariates, in which the point parameters of interest are  the point effects  of    treatments considered as single-point treatments. We (1)
identify    net effects of treatments  by    point effects of treatments, (2) express patterns of   net effects of treatments  by constraints on   point effects of treatments, and (3) show that all   sequential causal effects are determined by   net effects of treatments. Accordingly we (1) estimate  net effects of treatments through point effects of treatments by maximum likelihood, (2) improve the estimation by constraints on point effects of treatments and   assignment conditions of   treatments, and  (3)  use the estimates of   net effects of treatments to obtain those of   sequential causal effects.
As a result,
we obtain unbiased consistent maximum-likelihood   estimates of  sequential causal effects even
for   long treatment sequences. For illustration of our method, we study the causal effects  of various sequences of recreational drugs on the CD4 count among HIV patients.
\end{abstract}

\noindent%
\vfill

\newpage
\spacingset{1.45} 

\newtheorem{E}{Example}
\newtheorem{T}{Theorem}
\newtheorem{PP}{Proposition}
\newtheorem{C}{Corollary}
\newtheorem{D}{Definition}
\newtheorem{R}{Remark}
\newtheorem{LL}{Lemma}

\section{Introduction}
In many economic and medical practices,  treatments are assigned in the form of a sequence to influence an outcome of interest that occurs after the last treatment of the sequence. Between treatments  there often exist time-dependent covariates that   may be posttreatment variables of the earlier treatments (Rosenbaum 1984;
Robins 1989; Frangakis \& Rubin 2002) and confounders of the subsequent  treatments.
One wishes to infer the causal effect of the treatment sequence on  the outcome often called the sequential causal effect.

Consider  parametrization for the conditional distribution of an  outcome given all     treatments   and     time-dependent covariates. In  the standard parametrization,  one usually uses the means of the  outcome given all the treatments and   time-dependent covariates  as standard parameters.
Robins (1986, 1997, 1999, 2004, 2009)  illustrated that an unsaturated outcome model imposing equalities
between   standard parameters   leads to erroneous rejection
of the null hypothesis of sequential causal effects if the  time-dependent covariates are
simultaneously  posttreatment variables of the earlier treatments and confounders  of the subsequent  treatments.
As the treatment sequence gets long, the number of standard parameters becomes huge, and with no constraint on these parameters, the maximum-likelihood (ML)   estimates of sequential causal effects may not be consistent (Robins and Ritov 1997; Robins 1997).
To obtain non-genuine likelihood-based   estimates of sequential causal effects,
two  semi-parametric approaches   have been developed:  the structural nested model  (Robins 1992,
1997, 2004, 2009; Robins et al. 1999; Murphy 2003; Henderson et al. 2010) and the marginal structural model (Robins 1999, 2009;
Murphy et al. 2001).

In this article, instead of standard parametrization, we  construct a point
parametrization for the  conditional  distribution of  an  outcome given  all     treatments and   time-dependent  covariates  and  develop a parametric approach to   sequential causal inference.
In  Section $2$, we introduce the  background and notation of   parametric sequential
causal inference.
In Section $3$,
we construct a point
parametrization by using the point effects of treatments   as the point parameters of interest and analyze  relationship between the  point effects of treatments, the net effects of treatments and the sequential causal effects.
In Section $4$, we
 estimate  net effects of treatments and sequential causal effects through point effects of treatments  by maximum likelihood and improve the estimation by constraints on point effects of treatments and   assignment conditions of treatments.
 In Section $5$, we illustrate our method by   an analytical example, a simulation study and a real example. In Section $6$, we conclude the article
with remarks.

\section{Background and Notation}

\subsection{Sequential causal effects}

Let  $z_t$ indicate the treatments at time $t$ ($t=1,\ldots,T$). Assume that all  $z_t$ are   discrete variables  and take the values $0,1,\ldots$.
We take $z_t=0$ as control treatment and $z_t=1,2,\ldots$ as active treatments.
Let
$\mathbf{z}_1^t=(z_1,\ldots, z_t)$ indicate the treatment sequences
from times  $1$ to   $t$. Suppose that every treatment sequence $\mathbf{z}_1^T$ could be applied to each unit of a population.
Assume that there is no
interference between units and  no represented treatment sequence for
any unit.
 For notational simplicity, we use
 one subpopulation defined  by stationary covariates of the population as our population,  and henceforth do not consider  stationary covariates in the following development.

 In the framework of sequential causal inference,   each unit could have a potential (time-dependent)
covariate vector $\mathbf{x}_t(\mathbf{z}_1^{t})$   between treatments $z_t$ and $z_{t+1} $    and    a potential outcome
$y(\mathbf{z}_{1}^T)$ of our interest after last treatment $z_T$ under  treatment sequence
$\mathbf{z}_1^T$.
Assume that $\mathbf x_t(\mathbf{z}_1^{t})$ is a discrete  vector with non-negative components. We take $\mathbf x_t(\mathbf{z}_1^{t})=\mathbf 0$ as  reference level.
Let
$\mathbf{x}_{1}^{t}(\mathbf z_1^{t})$ $=$ $\{\mathbf{x}_{1}(z_1),
\mathbf{x}_{2}(\mathbf{z}_1^2),\ldots,
\mathbf{x}_{t}(\mathbf{z}_1^{t})\}$  be the potential
covariate array between treatments $z_1$ and $z_{t+1}$.

In the above description of treatment sequence, potential covariates and potential outcome, every   $z_t$  in    $\mathbf{z}_1^T$ is a deterministic function of the earlier
treatments and potential covariates,
i.e. $z_t=z_t\{\mathbf{z}_1^{t-1},\mathbf{x}_1^{t-1}(\mathbf{z}_1^{t-1})\}$. If each $
{z}_t$  in $\mathbf{z}_1^T$ does not depend on the earlier treatments and potential covariates,   then
 $\mathbf{z}_1^{T}$ is   a static treatment sequence, and otherwise, it is
a dynamic treatment sequence. In this article, we study
the  sequential causal effect
\begin{equation}\label{eq0}
{\rm sce}(\mathbf a_1^T, \mathbf b_1^T)=E\{y(\mathbf{z}_{1}^T=\mathbf  a_1^T)\}- E\{y(\mathbf{z}_{1}^T=\mathbf
b_1^T)\},
\end{equation}
which is the causal effect of treatment sequence $\mathbf a_1^T$ relative to treatment sequence $\mathbf b_1^T$,  where $E\{y(\mathbf{z}_{1}^T)\}$ is the mean of the potential outcome  $y(\mathbf{z}_{1}^T)$ of the population under treatment sequence $\mathbf{z}_1^T$.

\subsection{Treatment assignment  and the $G$-formula}
In most practices,  treatments  ${z}_t$ ($t=1,\ldots, T$) are consecutively and randomly assigned, so  the potential
covariate vectors $\mathbf{x}_t(\mathbf{z}_1^{t})$ ($t=1,\ldots, T-1$) and the potential outcome
$y(\mathbf{z}_{1}^T)$ become consecutively and randomly observable. Denote the
observable    covariate vector simply by $\mathbf x_t$ ($t=1,\ldots, T-1$) and the observable outcome by $y$.
Let
$\mathbf{x}_1^{t}=(\mathbf{x}_1,\ldots, \mathbf{x}_{t})$. In order to use the observable variables $(\mathbf z_1^T, \mathbf{x}_1^{T-1}, y ) $ to identify the sequential causal effect ${\rm sce}(\mathbf a_1^T, \mathbf b_1^T)$, some assumptions are needed.

The randomly assigned treatments $\mathbf{z}_1^{t-1}$ are assumed to be equivalent to a deterministic treatment sequence, static or dynamic, which leads to the observable   covariate array  $\mathbf x_1^{t-1}$, and this assumption is  known as the consistency assumption (Robins, 1986, 1989, 1992, 1997, 1999, 2004, 2009). Denote the set of these observable variables in such a relationship  by $ (\mathbf z_1^{t-1},\mathbf{x}_1^{t-1})$.
Let $\mathbf z_t^{T}=(z_t,\ldots, z_{T})$ be the treatment sequence given   $(\mathbf z_1^{t-1}, \mathbf x_1^{t-1})$. Under  $\mathbf
z_t^{T}$ given $(\mathbf z_1^{t-1}, \mathbf x_1^{t-1})$, each unit could have  potential
covariate vectors $\mathbf x_t( z_1^{t})$, $\ldots$, $\mathbf
x_{T-1}(\mathbf z_1^{T-1})$  and a potential outcome $y(\mathbf z_1^T)$. Let $\mathbf x_t^{T-1}(\mathbf z_t^{T-1})=\{\mathbf x_t(\mathbf z_1^{t}),\ldots, \mathbf x_{T-1}(\mathbf z_1^{T-1})\}$ and $y(\mathbf z_t^{T})=y(\mathbf z_1^{T})$ for given $(\mathbf z_1^{t-1}, \mathbf x_1^{t-1})$.

In the following assumption, let   $z^{*}_t$  indicate  the treatments randomly assigned at time $t$.
The assignment of
$z^{*}_t$   ($t= 1, \ldots, T$) is assumed to satisfy
\begin{equation}\label{eq1}
\left\{ \begin{array}{l} \mathbf x_t^{T-1}(\mathbf z_t^{T-1}),
y(\mathbf z_t^{T})  \bot z^{*}_t \mid \mathbf z_1^{t-1}, \mathbf
x_1^{t-1} \\
0 < {\rm pr}(z^{*}_t\mid \mathbf z_1^{t-1},\mathbf x_1^{t-1}) < 1
\end{array} \right.
\end{equation}
for any   treatment sequence  $\mathbf z_t^T$   given the
observable variables $(\mathbf z_1^{t-1}, \mathbf x_1^{t-1})$.
Here
$A  \bot  B \mid C$ means that $A$ is conditionally independent of
$B$ given $C$,  and noticeably,  $\mathbf z_t^T$ is a deterministic treatment sequence  whereas $z^{*}_t$,  $\mathbf x_t^{T-1}(\mathbf z_t^{T-1})$ and $y(\mathbf z_t^{T})$ are random variables.
The first part of  (\ref{eq1}) is known as
the
assumption of no unmeasured confounders (Robins 1986, 1989, 1992, 1997, 1999, 2004, 2009). The  second part is known as the positivity assumption.
  There may exist other observable   covariates than
$\mathbf{x}_1^{T-1}$    but   the  assignment   of $z_t^*$ does not depend on
them and no further information is available about them.

Throughout the article, we adopt the following  notational conventions. First, the notations
$\mathbf z_{u}^{v}$, $\mathbf x_{u}^{v}$ and $\mathbf
x_u^{v}(\mathbf z_u^{v})$ with $u > v$ or $u=v=0$ or both $u<0$ and $v<0$  should be omitted from    relevant
expressions.
Thus,  the notations $\mathbf z_1^{0}$ and $\mathbf x_1^{0}$
in (\ref{eq1}) for $t=1$ should be omitted, and then (\ref{eq1}) is
 $ \mathbf
x_1^{T-1}(\mathbf z_1^{T-1}), y(\mathbf z_1^T)  \bot z^{*}_1$ and $0
< {\rm pr}(z^{*}_1) < 1$. Similarly, the notation
$ \mathbf
x_T^{T-1}(\mathbf z_T^{T-1})$ in  (\ref{eq1}) for $t=T$  should be omitted, and then
(\ref{eq1}) is  $y(z_T)  \bot z^{*}_T \mid \mathbf z_1^{T-1}, \mathbf
x_1^{T-1} $ and $0
< {\rm pr}(z^{*}_T\mid \mathbf z_1^{T-1}, \mathbf
x_1^{T-1}) < 1$.   Second, the sigma notation  $\sum_{i=u}^v a_i$ with $v<u$ should be omitted from   relevant expression.
Third, the notations $\mathbf z_{u}^{v}$, $\mathbf x_{u}^{v}$, $\mathbf
x_u^{v}(\mathbf z_u^{v})$  and $\sum_{i=u}^v a_i$  with $u < 1 $ and $v \geq 1$ are treated  as $\mathbf z_{1}^{v}$, $\mathbf x_{1}^{v}$, $\mathbf
x_1^{v}(\mathbf z_1^{v})$ and $\sum_{i=1}^v a_i$. Fourth, the notation $(\mathbf z_u^{v},\mathbf x_u^{v-1})$ is equal to $(\mathbf z_u^{v-1}, \mathbf {x}_u^{v-1}, z_v)$, and $(\mathbf z_u^{v},\mathbf x_u^{v})$ to $(\mathbf z_u^{v}, \mathbf {x}_u^{v-1}, \mathbf x_v)$; we may use one  or another notation in different contexts.

Standard parameters for the conditional distribution of  the observable outcome $y$ given   $(\mathbf z_1^{T},\mathbf {x}_1^{T-1})$  are the means $E(y \mid
  \mathbf z_1^{T},\mathbf {x}_1^{T-1})$, denoted by $\mu(\mathbf z_1^{T},\mathbf {x}_1^{T-1})$.  Standard parameters for the conditional distribution of the observable covariate $\mathbf x_t$ given the observable variables $(\mathbf z_1^{t},\mathbf {x}_1^{t-1})$ are the probabilities ${\rm pr}(\mathbf x_t\mid \mathbf z_1^{t},\mathbf
x_1^{t-1})$.
Using  assumption (\ref{eq1}), Robins (1986,
1997) derived the
well-known $G$-computation algorithm formula (also called $G$-formula)
\begin{equation}\label{eq8}
E\{y(\mathbf z_{1}^T)\}=\sum_{\mathbf x_1^{T-1}} \mu(\mathbf z_1^{T},\mathbf x_1^{T-1})
\prod_{t=1}^{T-1} {\rm pr}(\mathbf x_t\mid \mathbf z_1^{t},\mathbf
x_1^{t-1}),
\end{equation}
where treatment sequence $\mathbf z_{1}^T$ can be static or dynamic. According to (\ref{eq0}), we then have the $G$-formula for the sequential causal effect
$$
{\rm sce}(\mathbf a_1^T, \mathbf b_1^T)=
$$
$$
\sum_{\mathbf x_1^{T-1}} \mu(\mathbf a_1^{T},\mathbf x_1^{T-1})
\prod_{t=1}^{T-1} {\rm pr}(\mathbf x_t\mid \mathbf a_1^{t},\mathbf
x_1^{t-1})-\sum_{\mathbf x_1^{T-1}} \mu(\mathbf b_1^{T},\mathbf x_1^{T-1})
\prod_{t=1}^{T-1} {\rm pr}(\mathbf x_t\mid \mathbf b_1^{t},\mathbf
x_1^{t-1}),
$$
which identifies   sequential causal effect  by   standard parameters
 under assumption (\ref{eq1}).

\subsection{Conditional  distribution of observable outcome}

Instead of one set $(\mathbf z_{1}^T, \mathbf{x}_{1}^{T-1},y )$ of the observable variables,  we consider
 $N$ independent and identically distributed  sets,
 $\{
\mathbf z_{i1}^T, \mathbf{x}_{i1}^{T-1}, y_i\}$, $i=1,\ldots, N$.
The $G$-formula above  implies that  in parametric inference of  ${\rm sce}(\mathbf a_1^T, \mathbf b_1^T)$, we need to parameterize the following two distributions
$$
\prod_{i=1}^N \prod_{t=1}^{T-1}f(\mathbf x_{it}\mid \mathbf z_{i1}^{t},\mathbf
x_{i1}^{t-1})
$$
and
\begin{equation}\label{eq8_a}
\prod_{i=1}^N f(y_i\mid \mathbf z_{i1}^T, \mathbf{x}_{i1}^{T-1} ),
\end{equation}
where    $f(u\mid v)$ is a conditional probability distribution of $u$ given $v$ if   $u$ is discrete, or a conditional density distribution of $u$ given $v$ if $u$ is continuous.

 If $\mathbf x_t$ ($t=1,\ldots, T-1$) are   posttreatment variables of   $z_s$ ($s \leq t $), then the standard parameters  ${\rm pr}(\mathbf x_t\mid \mathbf z_1^{t},\mathbf
x_1^{t-1})$ and $\mu(\mathbf z_1^{T},\mathbf {x}_1^{T-1})$  essentially do not have any  pattern (Rosenbaum 1984;
Robins 1989; Frangakis \& Rubin 2002). If $\mathbf x_t$  are  simultaneously confounders of   $z_s$ ($s \geq t+1 $), then one needs to use all  these standard parameters to identify  ${\rm sce}(\mathbf a_1^T, \mathbf b_1^T)$.
With a long treatment sequence, the number of these parameters is huge. Without constraints on these parameters,     the  ML estimate of  ${\rm sce}(\mathbf a_1^T, \mathbf b_1^T)$ may not be consistent (Robins 1986, 1997, 1999, 2004, 2009; Robins \& Ritov 1997).

In this article we focus on parametrization of (\ref{eq8_a}). Henceforth we ignore the variability of $\{
\mathbf z_{i1}^T, \mathbf{x}_{i1}^{T-1}\}_{i=1}^N$ and treat the proportions as the probabilities.
We are going to construct a point parametrization for (\ref{eq8_a})  and use the point parameters to infer ${\rm sce}(\mathbf a_1^T, \mathbf b_1^T)$.

\section{Point Parametrization and Parametric Sequential
Causal Inference}

\subsection{Point  parametrization}
Given   $N$ sets $\{\mathbf z_{i1}^T, \mathbf x_{i1}^{T-1}\}_{i=1}^N$ of treatments and covariates, a stratum is a set of those   sets   satisfying certain condition. For instance, stratum $(\mathbf z_1^{t}, \mathbf x_1^{t-1})$ is a set of those   sets    satisfying
$(\mathbf z_{i1}^{t}, \mathbf x_{i1}^{t-1})=(\mathbf z_1^{t}, \mathbf x_1^{t-1})$.
Let   ${\rm pr}(A)$ denote the proportion of stratum $A$ in the $N$ sets and ${\rm pr}(A\mid B)$ denote the conditional proportion of stratum $A$ in stratum $B$.

Consider the mean   of  $y$ in stratum $(\mathbf z_1^{t}, \mathbf x_1^{t-1})$
\begin{equation}\label{nn1_0}
\mu(\mathbf z_1^{t}, \mathbf x_1^{t-1})= \sum_{\mathbf z_{t+1}^{T},
\mathbf
x_{t}^{T-1}} \mu( \mathbf z_1^{T},\mathbf
x_1^{T-1}) {\rm pr}(\mathbf z_{t+1}^{T},\mathbf x_{t}^{T-1}\mid
\mathbf z_1^{t},\mathbf x_1^{t-1})
\end{equation}
for $t=1,\ldots, T-1$ and  $\mu(\mathbf z_1^{T}, \mathbf x_1^{T-1})$.
The point effect of treatment $z_t>0$ on  stratum
$(\mathbf z_1^{t-1}, \mathbf x_1^{t-1})$ is then
\begin{equation}\label{nn1}
\theta(\mathbf z_1^{t-1}, \mathbf x_1^{t-1}, z_t)=\mu(\mathbf
z_1^{t-1}, \mathbf x_1^{t-1}, z_t ) -\mu(\mathbf z_1^{t-1}, \mathbf
x_1^{t-1}, z_t=0),
\end{equation}
where $\mu(\mathbf
z_1^{t-1}, \mathbf x_1^{t-1}, z_t )=\mu(\mathbf
z_1^{t}, \mathbf x_1^{t-1} )$ according to the notational convention given  in Section $2.2$.

Consider
the mean   of   $y$ in stratum $(\mathbf z_1^{t}, \mathbf x_1^{t})$
\begin{equation}\label{nn2_0}
\mu(\mathbf z_1^{t}, \mathbf x_1^{t})= \sum_{\mathbf z_{t+1}^{T},
\mathbf
x_{t+1}^{T-1}} \mu( \mathbf z_1^{T},\mathbf
x_1^{T-1}) {\rm pr}(\mathbf z_{t+1}^{T},\mathbf x_{t+1}^{T-1}\mid
\mathbf z_1^{t},\mathbf x_1^{t})
\end{equation}
for $t=1,\ldots,T-1$.
The point effect of   covariate   $\mathbf
x_{t}> \mathbf 0$ on   stratum $(\mathbf z_1^{t}, \mathbf x_1^{t-1})$ is then
\begin{equation}\label{nn2}
\gamma(\mathbf z_1^{t}, \mathbf x_1^{t-1},\mathbf x_{t})=\mu(\mathbf
z_1^{t}, \mathbf x_1^{t-1}, \mathbf x_{t}  )-\mu(\mathbf z_1^{t},
\mathbf x_1^{t-1}, \mathbf x_{t}=\mathbf 0).
\end{equation}

The grand mean is
\begin{equation}\label{na11_0}
\mu=\sum_{\mathbf z_1^{T},\mathbf x_1^{T-1}}\mu(\mathbf
z_1^{T},\mathbf x_1^{T-1}){\rm pr}(\mathbf z_1^{T},\mathbf
x_1^{T-1}).
\end{equation}
Given $\{
\mathbf z_{i1}^T, \mathbf{x}_{i1}^{T-1}\}_{i=1}^N$,  then  $ \theta(\mathbf z_1^{t-1},\mathbf x_1^{t-1}, z_t)$   ($t=1,\ldots, T$),    $\gamma(\mathbf z_1^{t},\mathbf x_1^{t-1},\mathbf x_t )$   ($t=1,\ldots, T-1$) and $\mu$ are parameters  for (\ref{eq8_a}), which are called
point parameters.

 From (\ref{nn1_0}-\ref{na11_0}), we see that each point parameter can be expressed in terms of the standard parameters
$\mu(\mathbf z_1^T,\mathbf x_1^{T-1})$.
Conversely,
we show in  Supplementary Material A that each standard parameter  can be expressed in terms of the point parameters by
$$
\mu(\mathbf z_1^T,\mathbf x_1^{T-1}) =\sum_{t=1}^T \left
[\sum_{z_t^*} - \theta(\mathbf z_1^{t-1},\mathbf x_1^{t-1},
z_t^*){\rm pr}(z_t^*\mid \mathbf z_1^{t-1},\mathbf x_1^{t-1}) +
\theta(\mathbf z_1^{t-1},\mathbf x_1^{t-1}, z_t)\right ]+
$$
\begin{equation}\label{na11}
\sum_{t=1}^{T-1} \left [\sum_{\mathbf x_{t}^*}
-\gamma(\mathbf z_1^{t},\mathbf x_1^{t-1}, \mathbf x_{t}^*){\rm
pr}(\mathbf x_{t}^*\mid \mathbf z_1^{t},\mathbf x_1^{t-1})+
\gamma(\mathbf z_1^{t},\mathbf x_1^{t-1}, \mathbf x_{t})\right ]+
\mu.
\end{equation}
Here we take $\theta(\mathbf z_1^{t-1},\mathbf x_1^{t-1}, z_t=0)=0$ and $\gamma(\mathbf z_1^{t},\mathbf x_1^{t-1}, \mathbf x_{t}=\mathbf 0)=0$.
Let $\Psi=\{\theta(\mathbf z_1^{t-1},\mathbf x_1^{t-1}, z_t), t=1,\ldots, T; \gamma(\mathbf z_1^{t},\mathbf x_1^{t-1}, \mathbf x_{t}), t=1,\ldots, T-1; \mu\}$ be the set of all point parameters. Then $\Psi$  forms  a new parametrization   of (\ref{eq8_a}), which is called  \textbf {point parametrization}.

\subsection{Net  versus point effects of treatments}
 The  net effect  of treatment $z_t>0$ on  stratum $(\mathbf z_1^{t-1}, \mathbf
x_1^{t-1})$  is
\begin{equation}\label{n_1_0_0}
\phi(\mathbf z_1^{t-1}, \mathbf
x_1^{t-1},z_t) =
\end{equation}
$$
E\{y( z_t, \mathbf z_{t+1}^{T}=\mathbf 0)\mid
\mathbf z_1^{t-1}, \mathbf x_1^{t-1}\}- E\{y( z_t =0, \mathbf
z_{t+1}^{T}=\mathbf 0)\mid \mathbf z_1^{t-1}, \mathbf x_1^{t-1}\},
$$
which is the  causal  effect of treatment sequence   $(z_t>0, \mathbf z_{t+1}^T=\mathbf 0)$ on    stratum $(\mathbf z_1^{t-1},
\mathbf x_1^{t-1})$ (Robins 1992, 1997, 1999, 2004, 2009).
The net effect   is also called the
blip effect     in the context of semi-parametric sequential causal inference. By constructions of (\ref{n_1_0_0}) and (\ref{nn1}), there are as many net effects of treatments as point effects of treatments. We are going to derive  relationship between the net and point effects.

Using  assumption (\ref{eq1}) and formula  (\ref{nn1_0}),
 we
 express, in  Supplementary Material A,   the mean $\mu(\mathbf z_1^{t},\mathbf x_1^{t-1} )$ in terms of the net effects since time $t$
  by
$$
\mu(\mathbf z_1^{t},\mathbf x_1^{t-1} )=E\{y(\mathbf z_t^T=\mathbf 0
)\mid \mathbf z_1^{t-1},\mathbf x_1^{t-1}\}+\phi(\mathbf z_1^{t-1},\mathbf x_1^{t-1},z_t ) \ +
$$
\begin{equation}\label{neq1_0}
 \sum_{s=t+1}^T\sum_{
\mathbf z_{t+1}^{s-1},\mathbf x_{t}^{s-1}}\sum_{z_s>0  }\phi(\mathbf z_1^{s-1}, \mathbf
x_1^{s-1},z_s){\rm pr}(
\mathbf z_{t+1}^{s-1},\mathbf x_{t}^{s-1},z_s\mid \mathbf
z_1^{t},\mathbf x_1^{t-1})
\end{equation}
for $t=1,\ldots, T-1$ and
  $$
\mu(\mathbf z_1^{T},\mathbf x_1^{T-1} )=E\{y( z_T=  0
)\mid \mathbf z_1^{T-1},\mathbf x_1^{T-1}\}+\phi(\mathbf z_1^{T-1},\mathbf x_1^{T-1},z_T )  .
$$
Formula (\ref{neq1_0}) implies that the mean $\mu(\mathbf z_1^{t},\mathbf x_1^{t-1} )$ arises from the net effects of   treatments since time $t$  on substrata   in stratum $(\mathbf z_1^{t},\mathbf x_1^{t-1})$.
This formula can also be derived from formula ($8.3$) of
Robins (1997).

Suppose that the data-generating mechanism implies a pattern of net effects:
there are only $K$ distinct net effects denoted by the net effect vector $\phi=(\phi_1,\ldots,\phi_K)$.
Accordingly, each $\phi_k$ corresponds to a set $S_k$  of strata    $(\mathbf z_1^{t-1},
\mathbf x_1^{t-1}, z_t>0)$  such that $\phi(\mathbf z_1^{t-1}, \mathbf
x_1^{t-1},z_t)=\phi_k$, namely,
$S_k=\{(\mathbf z_1^{t-1},
\mathbf x_1^{t-1}, z_t>0): \phi(\mathbf z_1^{t-1}, \mathbf
x_1^{t-1},z_t)=\phi_k\}$. We call $S_1,\ldots, S_K$ classes of strata and  $z_t$  an active treatment of class $k$ if $(\mathbf z_1^{t-1},
\mathbf x_1^{t-1}, z_t>0) \in S_k$.

Combining this pattern with (\ref{neq1_0}) and then
noticing that
$$
{\rm pr}(
\mathbf z_{t+1}^{s-1},\mathbf x_{t}^{s-1},z_s>0\mid \mathbf
z_1^{t},\mathbf x_1^{t-1})={\rm pr}(\mathbf z_1^{s-1}, \mathbf x_1^{s-1},z_s>0\mid \mathbf
z_1^{t},\mathbf x_1^{t-1}),
$$
we obtain
$$
\mu(\mathbf z_1^{t},\mathbf x_1^{t-1} )=E\{y(\mathbf z_t^T=\mathbf 0
)\mid \mathbf z_1^{t-1},\mathbf x_1^{t-1}\}+ \sum_{k=1}^K \phi_k I_{S_k}(\mathbf z_1^{t-1},\mathbf x_1^{t-1}, z_t ) \ +
$$
$$
\sum_{k=1}^K\sum_{s=t+1}^T\sum_{(\mathbf z_1^{s-1}, \mathbf x_1^{s-1},z_s>0) \in S_k }\phi_k {\rm pr}(\mathbf z_1^{s-1}, \mathbf x_1^{s-1},z_s>0\mid \mathbf
z_1^{t},\mathbf x_1^{t-1})
$$
for $t=1,\ldots, T-1$ and
$$
\mu(\mathbf z_1^{T},\mathbf x_1^{T-1} )=E\{y(z_T=\mathbf 0
)\mid \mathbf z_1^{T-1},\mathbf x_1^{T-1}\}+ \sum_{k=1}^K \phi_k I_{S_k}(\mathbf z_1^{T-1},\mathbf x_1^{T-1}, z_T ),
$$
where the indicator function
$I_A(b)$   takes   one if $b\in A$ and    zero otherwise.

Now
we consider the difference
\begin{equation}\label{neq1}
\mu(\mathbf z_1^{t-1},\mathbf x_1^{t-1}, z_t) -\mu(\mathbf
z_1^{t-1},\mathbf x_1^{t-1}, z_t=0)=\sum_{k=1}^K \phi_k
c^{(k)}(\mathbf z_1^{t-1},\mathbf x_1^{t-1},z_t)
\end{equation}
for all $(\mathbf z_1^{t-1},\mathbf x_1^{t-1},z_t>0)$ at
$t=1,\ldots,T$, where
\begin{equation}\label{neq1_1}
 c^{(k)}(\mathbf z_1^{t-1},\mathbf x_1^{t-1},z_t)
 =I_{S_k}(\mathbf z_1^{t-1},\mathbf x_1^{t-1}, z_t )+\sum_{s=t+1}^T\sum_{(\mathbf z_1^{s-1}, \mathbf x_1^{s-1},z_s>0) \in S_k }
\end{equation}
$$  \{{\rm pr}(\mathbf z_1^{s-1}, \mathbf x_1^{s-1},z_s>0\mid \mathbf
z_1^{t-1},\mathbf x_1^{t-1},z_t)- {\rm pr}(\mathbf z_1^{s-1}, \mathbf x_1^{s-1},z_s>0\mid \mathbf
z_1^{t-1},\mathbf x_1^{t-1},z_t=0) \}
$$
for $t=1,\ldots, T-1$ and
$$c^{(k)}(\mathbf z_1^{T-1},\mathbf x_1^{T-1},z_T)
 =I_{S_k}(\mathbf z_1^{T-1},\mathbf x_1^{T-1}, z_T ).
 $$
  The constant $c^{(k)}(\mathbf z_1^{t-1},\mathbf x_1^{t-1},z_t)$
describes  the  difference between  proportions of active
treatments of class $k$  at   $s=t, \ldots, T$ in    stratum
$(\mathbf z_1^{t-1},\mathbf x_1^{t-1}, z_t
> 0)$ versus   in  stratum $(\mathbf z_1^{t-1},\mathbf x_1^{t-1}, z_t
= 0)$.

Combining (\ref{nn1}) and (\ref{neq1}), we obtain the following   constraint on   point effects  of treatments
\begin{equation}\label{eq17}
\theta(\mathbf z_1^{t-1},\mathbf x_1^{t-1}, z_t)=\sum_{k=1}^K \phi_k
c^{(k)}(\mathbf z_1^{t-1},\mathbf x_1^{t-1},z_t)
\end{equation}
for all $(\mathbf z_1^{t-1},\mathbf x_1^{t-1},z_t>0)$ at $t=1,\ldots, T$, where   $ c^{(k)}(\mathbf z_1^{t-1},\mathbf x_1^{t-1},z_t)$   is given by (\ref{neq1_1}).
Constraint (\ref{eq17})
 decomposes the point effect  $\theta(\mathbf z_1^{t-1},\mathbf x_1^{t-1}, z_t)$ into the net effects of treatments since time $t$.
Under (\ref{eq17}), the    net effects  are identified by the point effects   because the  net effects   are fewer than  the  point effects.

\subsection{Sequential causal effects versus net effects of treatments}
Using formula (\ref{eq8}) and assumption (\ref{eq1}), we derive, in   Supplementary Material A,
\begin{equation}\label{n_1_0_0_1}
E\{y(\mathbf z_{1}^T)\}=E\{y(\mathbf z_{1}^T=\mathbf 0)\} + \phi(z_1)+ \sum_{t=2}^T\sum_{\mathbf x_1^{t-1}} \phi(\mathbf z_1^{t-1}, \mathbf
x_1^{t-1},z_t)
\prod_{s=1}^{t-1} {\rm pr}(\mathbf x_s\mid \mathbf z_1^{s},\mathbf
x_1^{s-1}).
\end{equation}
Using the pattern  $\phi=(\phi_1,\ldots, \phi_K)$ of $\phi(\mathbf z_1^{t-1}, \mathbf
x_1^{t-1},z_t)$, we obtain
\begin{equation}\label{n_1_0_0_2}
E\{y(\mathbf z_{1}^T)\}=E\{y(\mathbf z_{1}^T=\mathbf 0)\} +\sum_{k=1}^K \phi_k q^{(k)}(\mathbf z_{1}^T),
\end{equation}
where
\begin{equation}\label{n_1_0_0_2_1}
q^{(k)}(\mathbf z_{1}^T)=I_{S_k}(z_1)+\sum_{t=2}^T\sum_{ \mathbf x_1^{t-1} }I_{S_k}(\mathbf z_1^{t-1}, \mathbf
x_1^{t-1},z_t )
\prod_{s=1}^{t-1} {\rm pr}(\mathbf x_s\mid \mathbf z_1^{s},\mathbf
x_1^{s-1}),
\end{equation}
which is the sum of proportions of active treatments of class $k$ under treatment sequence $\mathbf z_{1}^T$.
  Combining (\ref{n_1_0_0_2}) with (\ref{eq0}), we obtain
\begin{equation}\label{n_1_0_0_3}
 {\rm sce}(\mathbf a_{1}^T,\mathbf b_{1}^T)=\sum_{k=1}^K \phi_k \{q^{(k)}(\mathbf a_{1}^T)- q^{(k)}(\mathbf b_{1}^T)\}.
\end{equation}
To estimate  ${\rm sce}(\mathbf a_{1}^T,\mathbf b_{1}^T)$, we  first estimate $\phi=(\phi_1,\ldots, \phi_K)$ through $\theta(\mathbf z_1^{t-1}, \mathbf x_1^{t-1}, z_t)$ by   (\ref{eq17}) and then use the estimate $\hat\phi=(\hat\phi_1,\ldots, \hat\phi_K)$ to obtain the estimate $\widehat{{\rm sce}}(\mathbf a_{1}^T,\mathbf b_{1}^T)$ by (\ref{n_1_0_0_3}).

\section {Estimating  Sequential Causal Effects by Maximum Likelihood}
\subsection{Likelihood of point parameters and outcome model}
The data  comprises independent observations $\{\mathbf z_{i1}^{T},
\mathbf x_{i1}^{T-1},y_i\}$ on units $i=1,\ldots,N$.
Using the conditional outcome distribution (\ref{eq8_a}), we obtain
the following likelihood of the point parameters
\begin{equation}\label{rev_1_a}
L\{\Psi; \{{y}_i\}_{i=1}^N|\{\mathbf z_{i1}^{T}, \mathbf x_{i1}^{T-1}\}_{i=1}^N\}
=\prod_{i=1}^N f\{y_i\mid \mathbf z_{i1}^T, \mathbf{x}_{i1}^{T-1};  \mu(\mathbf z_{i1}^T,\mathbf x_{i1}^{T-1})\},
\end{equation}
where $\Psi$ is the set of  point parameters constructed in Section $3.1$ and $\mu(\mathbf z_{i1}^{T},\mathbf x_{i1}^{T-1})=\mu(\mathbf z_{1}^{T}=\mathbf z_{i1}^{T},\mathbf x_{1}^{T-1}=\mathbf x_{i1}^{T-1})$   is   expressed  by (\ref{na11}) in terms of the point parameters in $\Psi$.
The outcome model is
\begin{equation}\label{rev_1}
\mu_i=\mu(\mathbf z_{i1}^{T},\mathbf x_{i1}^{T-1}),
\end{equation}
where $\mu_i=E(y_i\mid \mathbf z_{i1}^{T},\mathbf x_{i1}^{T-1})$ is the mean of $y_i$ given $(\mathbf z_{i1}^{T},\mathbf x_{i1}^{T-1})$.
The constraint on the point parameters is (\ref{eq17}).

\subsection{Outcome of normal distribution}
Suppose that  the outcome  $y$ is normally distributed.  For simplicity, we assume that $y$ has a known variance, say,
one, for any given $(\mathbf z_1^{T}, \mathbf x_1^{T-1})$.
Let  $s(A)$ be the set of units in  stratum $A$. With likelihood (\ref{rev_1_a}), the score function for the standard parameter $\mu(\mathbf z_1^{*T},\mathbf
x_1^{*(T-1)})$  is
$$
 U_{\mu(\mathbf z_1^{*T},\mathbf
x_1^{*(T-1)})}=\sum_{i\in s(\mathbf z_1^{*T},\mathbf
x_1^{*(T-1)})}\{y_i- \mu(\mathbf z_1^{*T},\mathbf x_1^{*(T-1)})\}.
$$
Using
the Chain Rule and (\ref{na11}), we obtain the score function for the point parameter $\theta(\mathbf
z_1^{t-1},\mathbf x_1^{t-1},z_t)$
$$
U_{\theta(\mathbf z_1^{t-1},\mathbf x_1^{t-1},z_t)}=\sum_{\mathbf
z_1^{*T},\mathbf x_1^{*(T-1)}}U_{\mu(\mathbf z_1^{*T},\mathbf
x_1^{*(T-1)}) }\frac{\partial \mu(\mathbf z_1^{*T},\mathbf
x_1^{*(T-1)}) }{
\partial \theta(\mathbf z_1^{t-1},\mathbf
x_1^{t-1},z_t)}.
$$
As proved in   Supplementary Material A, we have
\begin{T}\label{nt1}
The score function $U_{\theta(\mathbf z_1^{t-1},\mathbf
x_1^{t-1},z_t)}$
 depends  only on the point effects   $\theta(\mathbf
z_1^{t-1},\mathbf x_1^{t-1}, z_t^*)$   at time  $t$  if the   outcome $y$  is  normally  distributed and has the same known variance for all  given  $(\mathbf z_1^{T}, \mathbf x_1^{T-1})$.
Therefore  the estimate   $\hat \theta(\mathbf z_1^{t-1},\mathbf x_1^{t-1}, z_t)$ at time $t$
is   independent of the   estimates  of  point parameters at the other times.
\end{T}

Using the Chain Rule and
  constraint (\ref{eq17}), we obtain the following
score function for the net effect ${\phi}_k$    ($k=1,\ldots, K$)
$$
U_{\phi_k}=\sum_{t=1}^T\sum_{\mathbf z_1^{t-1},\mathbf x_1^{t-1},z_t}
 U_{\theta(\mathbf z_1^{t-1},\mathbf x_1^{t-1},z_t)}
c^{(k)}(\mathbf z_1^{t-1},\mathbf x_1^{t-1},z_t).
$$
This score function depends only on the net effect vector $\phi=(\phi_1,\ldots,\phi_K)$, because $
c^{(k)}(\mathbf z_1^{t-1},\mathbf x_1^{t-1},z_t)$ are constants,  and
$  U_{\theta(\mathbf z_1^{t-1},\mathbf x_1^{t-1},z_t)}$ depend only
on $\theta(\mathbf z_1^{t-1},\mathbf x_1^{t-1},z_t^*)$    according to
Theorem \ref{nt1}, which in turn
 depend only on $\phi$
under     constraint (\ref{eq17}). Let $\mathbf
U_{\phi}=(U_{\phi_1}, \ldots, U_{\phi_K})$. Then the system $\mathbf U_{\phi}=\mathbf 0$    contains   $K$ likelihood equations  involving the $K$-dimensional  vector $\phi$ only.  The solution to the system  is the ML
estimate $\hat\phi=(\hat\phi_1,\ldots,\hat\phi_K)$. The covariance matrix
${\rm cov}(\hat\phi)$ is obtained by using the corresponding information.

Alternatively, we can estimate the net effect vector $\phi$ in the following way.
First, we estimate the mean $\mu(\mathbf z_1^{t-1},\mathbf x_1^{t-1},
z_t)$. The estimate $
\hat \mu(\mathbf z_1^{t-1},\mathbf x_1^{t-1},
z_t)$ is the average of $y$ in stratum $(\mathbf z_1^{t-1},\mathbf x_1^{t-1},
z_t)$.
Second,  we use $
\hat \mu(\mathbf z_1^{t-1},\mathbf x_1^{t-1},
z_t)$ to calculate  the estimate  $\hat\theta(\mathbf z_1^{t-1},\mathbf x_1^{t-1},z_t)$ according to (\ref{nn1}).
 Third,
 we
perform a linear regression of $\hat\theta(\mathbf
z_1^{t-1},\mathbf x_1^{t-1},z_t)$ on $c^{(k)}(\mathbf
z_1^{t-1},\mathbf x_1^{t-1},z_t)$ according to     constraint
(\ref{eq17}) to estimate $\phi$.

With the estimate $\hat\phi$,  we use
(\ref{n_1_0_0_3}) to calculate the estimate $\widehat{\rm sce}(\mathbf a_{1}^T,\mathbf b_{1}^T)$.  The procedure of estimating sequential causal effects will be further illustrated in Section $5$.
Clearly, the   estimate $\hat \mu(\mathbf z_1^{t-1},\mathbf x_1^{t-1},
z_t)$  is unbiased for finite sample. Thus $\hat\theta(\mathbf z_1^{t-1},\mathbf
x_1^{t-1},z_t)$ is unbiased.  Therefore $\hat\phi$ and $
\widehat  {\rm sce}(\mathbf a_{1}^T,\mathbf b_{1}^T)$ are unbiased.

Oftentimes, the dimension $K$ of the net effect vector   $\phi$ is   finite, that is,  the net effects $\phi(\mathbf z_1^{t-1}, \mathbf
x_1^{t-1},z_t)$   ($t=1,\ldots, T$) have a pattern of finite dimension.
According to (\ref{eq17}) treated as a regression model, the estimate $\hat \phi$ is consistent
if there exist at least $K$ different point effects  of treatments which contain the $K$-dimensional vector $\phi$ and whose estimates have  zero covariance matrices as the sample size $N$ tends to infinity. This condition can be satisfied even with long treatment sequences, for instance, if the treatment variable $z_t$ ($t=1,\ldots, T$) and the covariate vector $\mathbf x_t$ ($t=1,\ldots, T-1$) take  finite numbers of values.
Clearly, if $\hat\phi$ is consistent, so is $
\widehat  {\rm sce}(\mathbf a_{1}^T,\mathbf b_{1}^T)$.

\subsection{Outcome of normal distribution after a long treatment sequence}
 In   single-point causal inference, it is well known that treatment assignment conditions may reduce the number of parameters in  estimation of  the causal effect of a single-point treatment (Rosenbaum  \&
Rubin 1983; Rosenbaum 1995; Rubin 2005).
In sequential causal inference, we may also use  assignment conditions of individual treatments to reduce the number of point parameters in estimation of sequential causal effects.

For illustration,  we consider the   Markov process, a common   assignment mechanism of treatment sequence, in which  the
assignment of $z_t$ $(t=1,\ldots, T)$ depends only on a limited history of previous treatments and covariates, for instance, the latest treatment and covariate $(z_{t-1},\mathbf
x_{t-1})$. In this case, we have the  proportion equality
$$
{\rm pr}(\mathbf z_1^{t-2},\mathbf
x_1^{t-2}\mid z_{t-1}, \mathbf
x_{t-1},z_t)={\rm pr}(\mathbf z_1^{t-2},\mathbf
x_1^{t-2}\mid
z_{t-1},  \mathbf x_{t-1}),
$$
or equivalently,
$$
{\rm pr}(\mathbf z_1^{t-2},\mathbf
x_1^{t-2}\mid z_{t-1}, \mathbf
x_{t-1},z_t>0)={\rm pr}(\mathbf z_1^{t-2},\mathbf
x_1^{t-2}\mid
z_{t-1},  \mathbf x_{t-1},z_t=0).
$$
 Given a finite sample, we can only achieve approximate proportion equality. With different sample sizes, we have different levels of approximation, which are required for different studies.
For instance,   with a small sample, we can achieve  approximately the same marginal distribution of each variable of $(\mathbf z_1^{t-2},\mathbf x_1^{t-2})$  in stratum $(z_{t-1},\mathbf x_{t-1},z_t>0)$ versus stratum $(z_{t-1},$ $\mathbf x_{t-1},z_t=0)$, which is sufficient in some studies.

Under the proportion equality above,
the mean of  $y$ in stratum
$(z_{t-1}, \mathbf x_{t-1}, z_t)$ is then
$$
\mu(z_{t-1}, \mathbf x_{t-1}, z_t)= \sum_{\mathbf z_1^{t-2},\mathbf
x_1^{t-2}} \mu( \mathbf z_1^{t},\mathbf
x_1^{t-1}) {\rm pr}(\mathbf z_1^{t-2},\mathbf
x_1^{t-2}\mid z_{t-1},
\mathbf x_{t-1}, z_t)
$$
$$
=\sum_{\mathbf z_1^{t-2},\mathbf
x_1^{t-2}} \mu( \mathbf z_1^{t},\mathbf
x_1^{t-1}) {\rm pr}(\mathbf z_1^{t-2},\mathbf
x_1^{t-2}\mid z_{t-1},
\mathbf x_{t-1}).
$$
The  point effect of treatment $z_t>0$ on stratum $(z_{t-1}, \mathbf x_{t-1})$ is
\begin{equation}\label{eq5_3_1_0_0}
\theta(z_{t-1}, \mathbf x_{t-1}, z_t)=\mu(z_{t-1},\mathbf x_{t-1}, z_t)-\mu(z_{t-1}, \mathbf x_{t-1}, z_t=0).
\end{equation}
Stratum $(z_{t-1}, \mathbf x_{t-1})$  is much larger than stratum $(\mathbf z_1^{t-1},\mathbf
x_1^{t-1})$  and thus has a large  probability  of  having  both   active and control treatments of  $z_t$. Therefore  $\theta(z_{t-1}, \mathbf x_{t-1}, z_t)$   is estimable at large $t$ even with a small sample.
Averaging both   sides of (\ref{nn1}) with respect to ${\rm pr}(\mathbf z_1^{t-2},\mathbf
x_1^{t-2}\mid
z_{t-1}, \mathbf x_{t-1})$ and then using (\ref{eq5_3_1_0_0}), we obtain
\begin{equation}\label{eq5_3_1_0}
\theta(z_{t-1}, \mathbf x_{t-1}, z_t)=\sum_{\mathbf z_1^{t-2},\mathbf
x_1^{t-2}}\theta( \mathbf z_1^{t-1},\mathbf
x_1^{t-1},z_t){\rm pr}(\mathbf z_1^{t-2},\mathbf
x_1^{t-2}\mid
z_{t-1}, \mathbf x_{t-1}).
\end{equation}

Consider  a pattern $\phi=(\phi_1,\ldots, \phi_K)$ of   net effects
 such that treatments $z_t>0$ have the same net effect on  strata $(\mathbf z_{1}^{t-1}, \mathbf
x_{1}^{t-1})$ with the same last variables $(z_{t-1}, \mathbf
x_{t-1})$, namely, all strata
$(\mathbf z_{1}^{t-1}, \mathbf x_{1}^{t-1}, z_t>0)$ with the same
 $( z_{t-1}, \mathbf
x_{t-1},z_t>0)$   are in the same class.
This assumption is   testable by using (\ref{eq17}).  On the other hand, there is little chance to reject it for a finite sample and a long treatment sequence. To justify it,  we  should also take subject knowledge  into account.

Averaging  both   sides of   (\ref{eq17}) with respect to ${\rm pr}(
\mathbf z_{1}^{t-2},\mathbf x_{1}^{t-2}\mid  z_{t-1}, \mathbf x_{t-1})$ and using the pattern and (\ref{eq5_3_1_0}),  we  obtain  the following   constraint on point effects  of treatments
\begin{equation}\label{eq18_0}
\theta( z_{t-1},\mathbf x_{t-1},z_t)=\sum_{k=1}^K
\phi_k  c^{(k)}( z_{t-1},\mathbf
x_{t-1},z_t)
\end{equation}
for all $( z_{t-1},\mathbf x_{t-1},z_t>0)$ at   $t=1,\ldots, T$, where
$$
 c^{(k)}( z_{t-1},\mathbf x_{t-1},z_t)
  =I_{S_k}(z_{t-1},\mathbf x_{t-1},z_t)+\sum_{s=t+1}^T\sum_{( z_{s-1},\mathbf x_{s-1}, z_s>0) \in S_k }
 $$
$$
 \{{\rm pr}( z_{s-1},\mathbf x_{s-1}, z_s>0\mid  z_{t-1},\mathbf x_{t-1},z_t)
 -
 {\rm pr}( z_{s-1},\mathbf x_{s-1}, z_s>0\mid  z_{t-1},\mathbf x_{t-1},z_t=0) \}
$$
for $t=1,\ldots, T-1$ and
$$
 c^{(k)}(z_{T-1},\mathbf x_{T-1},z_T)=I_{S_k}( z_{T-1},\mathbf x_{T-1},z_T).
 $$
Constraint  (\ref{eq18_0})
 decomposes the point effect
$\theta(z_{t-1}, \mathbf x_{t-1}, z_t)$   into the net effects $\phi_1,\ldots, \phi_K$ of treatments since time $t$.

Theorem \ref{nt1} and formula (\ref{eq5_3_1_0}) imply that the estimate  $\hat\theta(z_{t-1}, \mathbf x_{t-1}, z_t)$ at time $t$ is independent of the estimates of point parameters at the other times, i.e. $\theta(\mathbf z_1^{s-1},\mathbf x_1^{s-1},  z_{s})$ including
$\theta(z_{s-1}, \mathbf x_{s-1}, z_s)$ with $t\neq s$,   $\gamma(\mathbf z_1^{s},\mathbf x_1^{s-1},\mathbf x_{s})$ ($s=1,\ldots, T-1$) and the grand mean  $\mu$.

All arguments and statements about the estimation procedure, unbiasedness and consistency of the ML estimates of the net effect vector
$\phi$ and the sequential causal effect ${\rm sce}(\mathbf a_{1}^T,\mathbf b_{1}^T)$ are carried over from those in the previous subsection, if we replace  $\mu(\mathbf z_1^{t-1},\mathbf x_1^{t-1},z_t)$ by
$ \mu(z_{t-1},\mathbf x_{t-1},z_t)$,   $\theta(\mathbf z_1^{t-1},\mathbf x_1^{t-1},z_t)$ by
$ \theta(z_{t-1},\mathbf x_{t-1},z_t)$,
$c^{(k)}(\mathbf
z_1^{t-1},\mathbf x_1^{t-1},z_t)$ by $c^{(k)}(
z_{t-1},\mathbf x_{t-1},z_t)$,
(\ref{nn1}) by (\ref{eq5_3_1_0_0}), and (\ref{eq17}) by (\ref{eq18_0}).

\subsection{Outcomes of other common distributions}
For outcomes of many common distributions,
the  estimate   $
\hat\mu(\mathbf z_1^{t-1},\mathbf x_1^{t-1}, z_t)$
 is also the average of $y$ in stratum $
(\mathbf z_1^{t-1},\mathbf x_1^{t-1}, z_t)$ and
$$\hat \theta(\mathbf
z_1^{t-1},\mathbf x_1^{t-1},z_t)=\hat \mu(\mathbf z_1^{t-1},\mathbf
x_1^{t-1}, z_t)-\hat \mu(\mathbf z_1^{t-1},\mathbf x_1^{t-1},
z_t=0),
$$
like outcome of normal distribution.  If the estimate $\hat \theta(\mathbf
z_1^{t-1},\mathbf x_1^{t-1},z_t)$ at time $t$ is independent of the estimates of point parameters at the other times, then we use the  method described in Sections $4.2$ and $4.3$ to estimate the net effect vector $\phi$ and the sequential causal effect $
 {\rm sce}(\mathbf a_{1}^T,\mathbf b_{1}^T)$.

For outcomes of some distributions such as the binomial one, it may happen that the estimate $\hat \theta(\mathbf
z_1^{t-1},\mathbf x_1^{t-1},z_t)$ at time $t$ is not independent of the estimates of point parameters at the other times.
On the other hand, the  estimates
$\hat
\mu(\mathbf z_1^{t-1},\mathbf x_1^{t-1},z_t)$ and thus $\hat \theta(\mathbf
z_1^{t-1},\mathbf x_1^{t-1},z_t)$    are highly robust to   point parameters at times  $s>t$ in most practical cases. Therefore $\hat \theta(\mathbf
z_1^{t-1},\mathbf x_1^{t-1},z_t)$ at time $t$ is weakly correlated with the estimates of point parameters at the other times, and   the correlation may be ignored.
Hence we can still  use the method described in Section $4.2$
to estimate $\phi$ and $
 {\rm sce}(\mathbf a_{1}^T,\mathbf b_{1}^T)$.
 The situation for $\hat \mu(
z_{t-1},\mathbf x_{t-1},z_t)$ and  $\hat \theta(
z_{t-1},\mathbf x_{t-1},z_t)$ under the Markov process is similar, and  we can use the  method described in Section $4.3$
to estimate $\phi$ and $
 {\rm sce}(\mathbf a_{1}^T,\mathbf b_{1}^T)$ for long treatment sequences.

The obtained estimates  $\hat\phi$ and $
\widehat  {\rm sce}(\mathbf a_{1}^T,\mathbf b_{1}^T)$  are    both unbiased and consistent, like those based on normal distribution.

\section{Illustration}
\subsection{Analytical example}
Consider a simple setting in which the treatment variables $z_t$ ($t=1,\ldots, T$) and the time-dependent covariates $x_{t}$ ($t=1,\ldots, T-1$) are all dichotomous.
Suppose a simple pattern of net
effects of treatments, in which all active treatments $z_t=1$   have the same net effect denoted by $\phi$.
Thus all strata  $(\mathbf z_1^{t-1},\mathbf x_1^{t-1}, z_t=1)$  belong to one class denoted by $S$.
Furthermore, suppose that the treatment assignment
mechanism is a Markov process in which the assignment of $z_t$ depends only on $(z_{t-1}, x_{t-1})$.

Given  $(z_{t-1}, x_{t-1})$,  there is only  one point effect $\theta( z_{t-1}, x_{t-1},  z_t=1)$  denoted  by $\theta(
z_{t-1}, x_{t-1})$;
in particular, $\theta(z_1=1)=\theta$ at $t=1$.
Similarly,   in (\ref{eq18_0}), denote $c^{(1)}(z_{t-1}, {x}_{t-1}, z_t=1)$ by $c(z_{t-1}, {x}_{t-1})$;
  in particular,  $c^{(1)}(z_1=1)=c$  at $t=1$. Then
  the  constraint (\ref{eq18_0})   becomes
\begin{equation}\label{n_6_0_0_0}
 \theta (z_{t-1}, {x}_{t-1})
=\phi c(z_{t-1}, {x}_{t-1})
\end{equation}
for $(z_{t-1}, {x}_{t-1})=(0,0),(0,1), (1,0), (1,1)$ at
  $t=1,\ldots, T$, where
$$
c(z_{t-1}, {x}_{t-1})=1+\sum_{s=t+1}^T \{ {\rm pr}
 (z_s=1\mid z_{t-1},x_{t-1},z_t=1)-{\rm
pr} (z_s=1\mid z_{t-1},x_{t-1},z_t=0)\}
$$
for $t=1,\ldots, T-1$ and $c(z_{T-1},x_{T-1})=1$.
Constraint (\ref{n_6_0_0_0})
 decomposes the point effect
$\theta(z_{t-1},   x_{t-1})$   into the net effect  $\phi$ of treatments since time $t$.

According to  (\ref{n_1_0_0_3}),
 the sequential causal effect is then
\begin{equation}\label{n_5_0_0_0}
 {\rm sce}(\mathbf a_{1}^T,\mathbf b_{1}^T)=\phi\{q(\mathbf a_1^T)-q(\mathbf b_1^T)\},
\end{equation}
where, according to (\ref{n_1_0_0_2_1}),
$$
q(\mathbf z_{1}^T)=I_{S}(z_1)+\sum_{t=2}^T\sum_{ \mathbf x_1^{t-1} }I_{S}(\mathbf z_1^{t-1}, \mathbf
x_1^{t-1},z_t )
\prod_{s=1}^{t-1} {\rm pr}(\mathbf x_s\mid \mathbf z_1^{s},\mathbf
x_1^{s-1}),
$$
 which is the sum of proportions of  active treatments in the treatment sequence. Noticeably, $q(\mathbf z_1^T)$ can be a non-integer if $\mathbf z_1^T $ is a dynamic treatment sequence.

Supposing that the outcome $y$ is normal with variance equal to one given $(\mathbf z_{1}^T,\mathbf x_{1}^{T-1})$, we estimate the sequential causal effect  ${\rm sce}(\mathbf a_{1}^T,\mathbf b_{1}^T)$ according to the procedure described in Section $4$. Let  $s(A)$ be the set of units in  stratum $A$ and $n(A)$ be the
number of units in   stratum $A$. In the first step, we calculate
$$
 \hat
\mu(z_{t-1}, x_{t-1}, z_t)
=\frac{\sum_{i\in s(z_{t-1}, x_{t-1}, z_t)}y_i}{n(z_{t-1},  x_{t-1}, z_t)},
$$
$$
{\rm var}\{ \hat
\mu(z_{t-1},  x_{t-1}, z_t)\}=\frac{1}{n(z_{t-1},  x_{t-1}, z_t)}.
$$
In the second step,  we  calculate by (\ref{eq5_3_1_0_0})
$$
\hat\theta(z_{t-1}, x_{t-1})=\hat\mu(z_{t-1},x_{t-1}, z_t=1)-\hat\mu(z_{t-1}, x_{t-1}, z_t=0),
$$
$$
{\rm var}\{ \hat\theta(z_{t-1},x_{t-1})\}={\rm var}\{ \hat
\mu(z_{t-1}, x_{t-1}, z_t=1)\}+{\rm var}\{ \hat
\mu(z_{t-1}, x_{t-1}, z_t=0)\}.
$$
In the third step, we treat constraint (\ref{n_6_0_0_0}) as regression to estimate the net effect $\phi$. In this simple regression, we first  estimate the net effect
 on stratum $(z_{t-1}, {x}_{t-1})$, i.e. $\phi(z_{t-1},x_{t-1})$, by
$$
\hat\phi(z_{t-1},x_{t-1})=\frac{\hat\theta (z_{t-1}, {x}_{t-1})}{c(z_{t-1}, {x}_{t-1})}
$$
$$
{\rm var}\{\hat\phi(z_{t-1},x_{t-1})\}=\frac{{\rm var}\{\hat \theta( z_{t-1}, x_{t-1})\}}{c^2(z_{t-1}, {x}_{t-1})},
 $$
 and then average  $\hat\phi(z_{t-1},x_{t-1})$ over all strata $(z_{t-1},x_{t-1})$ at   $t=1,\ldots, T$  to obtain
$$
\hat\phi=\frac{\sum_{t=1}^T\sum_{(z_{t-1},x_{t-1})}
\hat\phi(z_{t-1},x_{t-1})
/{\rm var}\{\hat\phi(z_{t-1},x_{t-1})\} }
{\sum_{t=1}^T\sum_{(z_{t-1},x_{t-1})} 1 /
{\rm var}\{\hat\phi(z_{t-1},x_{t-1})\}},
$$
$$
{\rm var}(\hat\phi)=\frac{1} { \sum_{t=1}^T\sum_{(z_{t-1},x_{t-1})}
1 / {\rm var}\{\hat\phi(z_{t-1},x_{t-1})\}}.
$$
In the last step,   we calculate  by (\ref{n_5_0_0_0})
$$
 \widehat{\rm sce}(\mathbf a_{1}^T,\mathbf b_{1}^T)=\hat\phi \{q( \mathbf a_1^T)-q(\ \mathbf b_1^T)\},
$$
$$
{\rm var}\{\widehat{\rm sce}(\mathbf a_{1}^T,\mathbf b_{1}^T)\}={\rm var}(\hat\phi) \{q( \mathbf a_1^T)-q( \mathbf b_1^T)\}^2.
$$
The obtained estimates $\hat\phi$ and $ \widehat{\rm sce}(\mathbf a_{1}^T,\mathbf b_{1}^T)$ are both unbiased and consistent.

It is theoretically possible to  do the same estimation in the  standard parametrization,  but practically difficult  due to high dimension of   standard parameters and complex expression of  constraint (\ref{n_6_0_0_0}) in terms of   standard parameters. Furthermore, if   equalities are imposed between standard parameters, then the  estimation is biased (Robins 1986, 1997, 1999, 2004, 2009).

\subsection{A simulation study}
Here we show by simulation that   interval estimation of  the sequential causal effect ${\rm sce}(\mathbf a_{1}^T,\mathbf b_{1}^T)$ achieves the nominal  coverage probability for the example of the previous subsection.  According to (\ref{n_5_0_0_0}), it is sufficient to study the net effect $\phi$.
We are going to simulate three situations with the net effect $\phi=-10,10, 0$ respectively. In particular, $\phi=0$ corresponds to the null hypothesis of ${\rm sce}(\mathbf a_{1}^T,\mathbf b_{1}^T)=0$ for all $\mathbf a_{1}^T$ and $\mathbf b_{1}^T$.
It is of considerable interest to simulate  a consistent estimation of the  causal effect of a treatment sequence of infinite length, but due to scope of this article, we only consider the case of  $T=3$.  Then the variables in temporal order are $(z_1, x_1, z_2, x_2, z_3, y)=(\mathbf z_1^3, \mathbf x_1^2, y)$.

We   construct  the standard parameters $\mu(\mathbf z_1^3, \mathbf x_1^2)$ which reflect the simple pattern of net effects, described in Section $5.1$, under which all net effects are the same and equal to $\phi$. A brief description of the procedure is given here  whereas a detailed description is presented in Table $1$.
First,
we construct the proportions of $z_1$, $x_1$, $z_2$, $x_2$, and $z_3$, which give a sample size  of $1232$ units yielding integer frequencies for all $(\mathbf z_1^3, \mathbf x_1^2)$.
Second, with the obtained proportions and a given value of  $\phi$, we calculate the point  effect $\theta(\mathbf z_1^{t-1},\mathbf x_1^{t-1}, z_t=1)$  ($t=1,2, 3$) according to constraint (\ref{eq17}). Denoting
$\theta(\mathbf z_1^{t-1},\mathbf x_1^{t-1},z_t=1)$ by $\theta(\mathbf z_1^{t-1},\mathbf x_1^{t-1})$ and $
c(\mathbf z_1^{t-1},\mathbf x_1^{t-1},z_t=1)$ by $c(\mathbf z_1^{t-1},\mathbf x_1^{t-1})$, constraint  (\ref{eq17}) is then
\begin{equation}\label{eq2015_1}
\theta(\mathbf z_1^{t-1},\mathbf x_1^{t-1})=\phi
c(\mathbf z_1^{t-1},\mathbf x_1^{t-1}),
\end{equation}
where
$$
c(\mathbf z_1^{t-1}, \mathbf {x}_1^{t-1})=1+\sum_{s=t+1}^T \{ {\rm pr}
 (z_s=1\mid \mathbf z_1^{t-1},\mathbf x_1^{t-1},z_t=1)-{\rm
pr} (z_s=1\mid \mathbf z_1^{t-1}, \mathbf x_1^{t-1},z_t=0)\}.
$$
The second part of the formula is used to calculate the constants
 $c(\mathbf z_1^{t-1},\mathbf x_1^{t-1})$   while the first part is used  to calculate  $\theta(\mathbf z_1^{t-1},\mathbf x_1^{t-1})$.
  Noticeably, $\theta(\mathbf z_1^{t-1},\mathbf x_1^{t-1})$ is not equal to $\theta( x_{t-1}, z_{t-1})$, and we use the former  to construct the standard parameter.
Third, we arbitrarily choose the point  effect $\gamma(\mathbf z_1^{t}, \mathbf x_1^{t})$   ($t=1,2$) and the grand mean $\mu$, which
according to Theorem \ref{nt1} do not affect the net effect.
Finally, we insert  the obtained point parameters $\theta(\mathbf z_1^{t-1},\mathbf x_1^{t-1})$, $\gamma(\mathbf z_1^{t}, \mathbf x_1^{t})$ and $\mu$ into (\ref{na11}) to obtain  the standard parameters $\mu(\mathbf z_1^3, \mathbf x_1^{2})$.
In this setting,  we have also made
 the assignment of $z_3$   depend only on  $(z_2, x_2)$ so that the treatment assignment is a Markov process.
Furthermore, the time-dependent covariate $x_1$ is a posttreatment variable of $z_1$ and a confounder of $z_2$ while the time-dependent covariate $x_2$ is a posttreatment variable of $z_2$ and a confounder of $z_3$.
The standard parameters and the relevant SAS code  are  given in  Supplementary Material B.

With the obtained standard parameter $\mu(\mathbf z_1^3, \mathbf x_1^{2})$, we  generate data.
Because we  ignore the sampling variability of treatments and covariates in this article, we only generate the outcome $y$ given $(\mathbf z_1^3, \mathbf x_1^2)$.
With  $\mu(\mathbf z_1^3, \mathbf x_1^{2})$, assuming that the variance of $y$ given $(\mathbf z_1^3, \mathbf x_1^2)$  is one, we generate $y$ to   form a data set of $1232$ observations on  $(\mathbf z_1^3, \mathbf x_1^2, y)$. A total of
$2000$ data sets are generated.

With the $2000$ data sets,  we calculate the actual coverage probability for the confidence interval of the net effect $\phi$.
For each   data set,
we  calculate the confidence interval of    $\phi$ as follows. Using the method described in the previous subsection,
we  first calculate the estimate  $\hat\theta(z_{t-1}, x_{t-1})$ and the constants $c(z_{t-1},x_{t-1})$   and then regress $\hat\theta(z_{t-1}, x_{t-1})$ on $c(z_{t-1}, x_{t-1})$  according to (\ref{n_6_0_0_0}) to  obtain the estimate $\hat\phi$. With $\hat\phi$  and its variance, we calculate the   confidence interval of $\phi$.
With $2000$ data sets, we obtain $2000$   confidence intervals.  By counting how many confidence intervals  contain the given value of $\phi$, we obtain  the actual  coverage probability for the   confidence interval of $\phi$.
The SAS code generating the data set and calculating the actual coverage probability is presented in Supplementary Material B. The mean and variance of     $\hat\phi$ and the actual coverage probability of the $95\%$ confidence interval of $\phi$  are presented in Table $2$ together with the given value of $\phi$.

Table $2$ shows that for the three simulations with $\phi=-10,10, 0 $ respectively,
the actual coverage probability for the $95\%$ confidence interval of $\phi$ is the same and equal to   $94.90$  \%.

\subsection{A real example}
\subsubsection{Medical background and the data}
Here we use a medical example to illustrate the practical   procedure of estimating sequential causal effects. Many HIV-infected patients use recreational drugs such as cocaine.
There are rich literatures about the immediate influence of the recreational drug  on
  the count of CD4 cells, which reflects progression of the disease.  Here we study the distant influence of the recreational drug on CD4 count when the drug is used repeatedly.

The Multicenter AIDS Cohort Study enrolled nearly $5000$ gay or bisexual men from Baltimore, Pittsburgh, Chicago and Los Angeles between $1984$ and $1991$, and required these men to  return   every 6 months to complete
a questionnaire and undergo various  examinations  (Kaslow et al. 1987).  Our data was a sub data of the study  which  involved
 $375$ participants who were seronegative at entry and seroconverted during the follow-up (Zeger \& Diggle 1994).
 In the initial period of the seroconversion, these participants were not exposed to anti-HIV drugs, which might complicate the sequential causal effects of  recreational drugs.
Hence we restricted our study to the visit before   seroconversion ($t=0$) and the first   and second   visits after   seroconversion ($t=1,2$).
Furthermore, some participants  lost their follow-ups  due to unknown non-ignorable missing data mechanism, so we excluded these participants and finally obtained a data of $256$ participants. The data set is presented in Supplementary Material B.

At each visit $t=0,1,2$, participants recalled or examined  the drug use ($z_t$), CD4 count ($x_{t1}$), the number of packs of cigarettes a day ($x_{t2}$), the number of sexual partners ($x_{t3}$) and a mental illness score ($x_{t4}$). We assumed that
 $z_t$  occurred prior to $x_{t1}, \ldots, x_{t4}$. Age   ($x_{05}$)  was also included as a covariate at visit $t=0$. Consequently, the temporal order of these variables is $\{z_0,  (x_{01},\ldots, x_{05}), z_1,$  $ (x_{11},\ldots, x_{14}), z_2, (x_{21},\ldots, x_{24})\}$.

The treatment variables are drug uses $z_1$ and $z_2$. Due to incomplete  information about covariates prior to drug use $z_0$, it is not possible to obtain the sequential causal effect concerning $z_0$. Instead, we use  $z_0$ and $(x_{01}, \ldots, x_{05})$ as stationary covariates in   adjustment of  participants' differences. The time-dependent covariates between $z_1$ and $z_2$ are $(x_{11},\ldots,x_{14})$.
The outcome is the logarithm of CD4 count  at $t=2$, i.e. $y={\rm log}(x_{21})$.
 All variables prior to $y$ or $x_{21}$ are dichotomized, with ones implying 'yes' or 'high' and zeros 'no' or 'low'.
  We assume that  the outcome $y$ is normally distributed. We wish to estimate the sequential  causal effect  of drug use $(z_1, z_2)$.

\subsubsection{Point and net effects of recreational drugs}
Let $\mathbf x_0=(z_0, x_{01}, x_{02},x_{03}, x_{04},x_{05})$, which is the stationary covariate vector, and $\mathbf x_1=(x_{11}, x_{12},x_{13}, x_{14})$, which is the time-dependent covariate vector between drug uses $z_1$ and $z_2$.
As described in Section $4$, we  use the  model, for $i=1,\ldots, 256$,
$$
\mu_i=\mu(\mathbf x_{i0},  z_{i1}, \mathbf x_{i1}, z_{i2})
$$
to estimate  the point effect $\theta(\mathbf x_0)$ of   $z_1=1$ and the point effect $\theta(\mathbf x_0,z_1,\mathbf x_1)$ of   $z_2=1$.
According to Theorem \ref{nt1}, the estimate $\hat\theta(\mathbf x_0)$  is independent of  $\hat \theta(\mathbf x_0,z_1,\mathbf x_1)$, so we can estimate the two point effects separately, as follows.
We (1) use the above  model
to estimate
the variance of $y$ given $(\mathbf x_0,z_1,\mathbf x_1, z_2)$, i.e. ${\rm var}(y\mid \mathbf x_0,z_1,\mathbf x_1, z_2) $, (2) use the same model to estimate  $\theta(\mathbf x_0,z_1,\mathbf x_1)$, and (3) use the model
$$
\mu_i=\mu(\mathbf x_{i0},  z_{i1})
$$
to estimate $\theta( \mathbf x_0)$   where the variance estimated from step (1) is used.

We improve the estimation in the usual framework of statistical modeling. By the likelihood ratio-based significance test of the model parameters at the significance level of $10$ \%,
we find that only the CD4 count  $x_{01}$  is significant,  so  $\theta( \mathbf x_0)$ is considered equal to the point effect $\theta(x_{01})$ of drug use $z_1=1$. Furthermore,
 $x_{01}$ does not have  interaction with $z_1$. Hence we use the model
\begin{equation}\label{eq40_b}
\mu_i=\mu( x_{i01}, z_{i1})=\alpha_1+z_{i1}\beta_1 + x_{i01}\gamma_{1}
\end{equation}
to estimate   $\beta_1$ which is equal to $\theta(x_{01})$.
Similarly, we find that only  the CD4 counts $x_{01}$ and  $x_{11}$  are significant, so $\theta(\mathbf x_0,z_1,\mathbf x_1)$ is considered equal to the point effect $\theta(x_{01}, x_{11})$ of drug use $z_2=1$. Furthermore,  $x_{01}$ and  $x_{11}$  do not have  interaction with $z_2$. Hence we use the model
\begin{equation}\label{eq40_c}
\mu_i=\mu( x_{i01}, x_{i11}, z_{i2})=\alpha_2+z_{i2}\beta_2+x_{i01}\gamma_{2}+ x_{i11}\gamma_{3}
\end{equation}
to estimate  $\beta_2$ which is equal to $\theta(x_{01}, x_{11})$.
In the estimation,  the variance of $y$ given $(  x_{01}, z_{1}, x_{11}, z_{2})$, i.e. ${\rm var}(y\mid   x_{01}, z_{1}, x_{11}, z_{2}) $, is  estimated  by using the  model
\begin{equation}\label{eq40_a}
\mu_i=\mu(  x_{i01}, z_{i1}, x_{i11}, z_{i2}).
\end{equation}
The estimates $\hat\beta_1$ and  $\hat\beta_2$ and their variances are presented in Table $3a$.

Now we estimate the net effect
 $\phi( x_{01})$ of drug use  $z_1=1$ and the net effect $\phi(x_{01},z_1,x_{11})$ of drug use  $z_2=1$.
 Given the  small sample,  it is reasonable to
assume a pattern of net effects
 $\phi( x_{01})=\phi_1$ for all $ x_{01}$  and   $\phi( x_{01},z_1,  x_{11})=\phi_2$ for all $( x_{01},z_1,  x_{11})$.
 To $\phi_1$ corresponds the class $S_1=\{(x_{01}, z_1=1)\}$. To $\phi_2$ corresponds the class $S_2=\{(x_{01}, z_1, x_{11}, z_2=1)\}$.

Because $\beta_1=\theta(x_{01})$ for all $ x_{01}$  and $\beta_2=\theta(x_{01}, x_{11})$ for all $( x_{01}, x_{11})$,
we  decompose the point effects $\beta_1$ and $\beta_2$  into the net effects $\phi_1$ and $\phi_2$  as
$$
 \beta_1=\phi_1 + \phi_2\{{\rm prop}(z_2=1\mid z_1=1)- {\rm prop}(z_2=1\mid z_1=0)\},
$$
\begin{equation}\label{eq40_d}
 \beta_2=\phi_2.
 \end{equation}
 According to this decomposition, we regress $\hat  \beta_1$ and $\hat  \beta_2$ on the proportions to obtain the estimates $\hat\phi_1$ and $\hat\phi_2$ and  their covariance matrix, which are presented in Table $3b$.

 In the procedure above, we estimated the variances  ${\rm var}(\hat\beta_1)$ and  ${\rm var}(\hat\beta_2)$ through ${\rm var}(y\mid  x_{01}, z_1, , x_{11},z_{2}) $ estimated from model (\ref{eq40_a}).
On the other hand, we can estimate  ${\rm var}(\hat\beta_1)$ through ${\rm var}(y\mid   x_{01}, z_1) $ estimated from model (\ref{eq40_b}), and ${\rm var}(\hat\beta_2)$  through ${\rm var}(y\mid  x_{01}, x_{11}, z_2) $  estimated from model
 (\ref{eq40_c}). But ${\rm var}(y\mid  x_{01}, z_1,  x_{11},z_{2})$ is smaller than ${\rm var}(y\mid  x_{01}, x_{11}, z_{2})$ and ${\rm var}(y\mid  x_{01}, z_1 )$, yielding smaller estimates for  ${\rm var}(\hat\beta_1)$ and  ${\rm var}(\hat\beta_2)$.

On the other hand,  it can be  difficult to estimate
the variance of the outcome $y$ given all the treatments and covariates  $(\mathbf z_{1}^{T},\mathbf x_{1}^{T-1})$, i.e.  ${\rm var}(y\mid \mathbf z_{1}^{T},\mathbf x_{1}^{T-1})$, particularly when treatment sequence is long.
As an illustration of possible way out, recall the example given in Section $5.1$, where we calculated the variance ${\rm var}\{\hat\theta(z_{t-1}, x_{t-1})\}$  through the variance ${\rm var}(y\mid \mathbf z_{1}^{T},\mathbf x_{1}^{T-1})$ assumed to be equal to one. If  ${\rm var}(y\mid \mathbf z_{1}^{T},\mathbf x_{1}^{T-1})$ is unknown and difficult to estimate,  we can estimate ${\rm var}\{\hat\theta(z_{t-1}, x_{t-1})\}$ through  ${\rm var}(y\mid  z_{t-1}, x_{t-1}, z_t)$ estimated from  the model
 $$
 \mu_i=\mu( z_{i(t-1)}, x_{i(t-1)}, z_{i(t)}).
 $$

 \subsubsection{Sequential causal effects}
 In this example, there are only two classes of strata:
$S_1=\{(x_{01}, z_1=1)\}$ and  $S_2=\{(x_{01}, z_1, x_{11}, z_2=1)\}$.
 Applying formulas (\ref{n_1_0_0_3}) and (\ref{n_1_0_0_2_1})    to each stratum defined by the stationary covariate  $x_{01}$ to obtain the   sequential causal effect on stratum $x_{01}$ and then  averaging the $x_{01}$-specific  sequential causal effect with respect to  ${\rm pr}( x_{01})$, we   obtain the sequential causal effect on the population
$$
{\rm sce}(\mathbf a_{1}^2,\mathbf b_{1}^2)=\phi_1 \{q^{(1)}(\mathbf a_{1}^2)- q^{(1)}(\mathbf b_{1}^2)\}
+
\phi_2 \{q^{(2)}(\mathbf a_{2}^2)- q^{(2)}(\mathbf b_{1}^2)\}
$$
where
$$
q^{(1)}(\mathbf z_{1}^2)=\sum_{x_{01}}I_{S_1}(x_{01},z_1) {\rm pr}(x_{01}),
$$
$$
q^{(2)}(\mathbf z_{1}^2)=\sum_{ x_{01}, x_{11} }I_{S_2}(x_{01},z_1,
x_{11},z_2 )
 {\rm pr}( x_{11}\mid  x_{01}, z_1){\rm pr}(x_{01}).
$$
Using the estimates $\hat\phi_1$ and $\hat\phi_2$ and their covariance matrix, we obtain the estimate $\widehat{\rm sce}(\mathbf a_{1}^2,\mathbf b_{1}^2)$.

To  illustrate use of the  formulas above,  we consider a sequential causal effect  with static treatment sequences $\mathbf b_{1}^2=(0,0)$ and $\mathbf a_{1}^2=(1,0)$.
For $\mathbf b_{1}^2=(0,0)$,   none of strata $(x_{01},z_1=0)$ belong to $S_1$ and nor strata $(x_{01},z_1=0,
x_{11},z_2=0)$ to  $S_2$,
so
we have $q^{(1)}(\mathbf b_{1}^2)=0$ and  $q^{(2)}(\mathbf b_{1}^2)=0$. For
 $\mathbf a_{1}^2=(1,0)$,  all strata $(x_{01},z_1=1)$ belong to $S_1$ whereas none of strata $(x_{01},z_1=1,
x_{11},z_2=0)$ belong to  $S_2$,
 so we have  $q^{(1)}(\mathbf a_{1}^2)=1$ and $q^{(2)}(\mathbf a_{1}^2)=0$. Hence we obtain  ${\rm sce}(\mathbf a_1^2, \mathbf b_1^2)=\phi_1$.

We also consider a sequential causal effect with one static treatment sequence   $\mathbf b_{1}^2=(0,0)$ and one dynamic  treatment sequence $\mathbf a_{1}^2$, in which   $a_1=1$ but $a_2=1$ when $x_{11}=0$ and $a_2=0$ when $x_{11}=1$.  For the dynamic treatment sequence, all strata $(x_{01},z_1=1)$ belong to $S_1$ whereas only strata  $(x_{01},z_1=1,
x_{11}=0,z_2=1)$ belong to  $S_2$, so
we have  $q^{(1)}(\mathbf a_{1}^2)=1$ and
$$
q^{(2)}(\mathbf a_{1}^2)=\sum_{x_{01}}{\rm pr}( x_{11}=0\mid  x_{01}, z_1=1){\rm pr}(x_{01}).
$$
Hence we obtain the sequential causal effect
$
{\rm sce}(\mathbf a_1^2, \mathbf b_1^2)=\phi_1+\phi_2 q^{(2)}(\mathbf a_{1}^2).
$

Table $3c$ presents estimates for a variety of sequential causal effects.
The first two sequential causal effects represent  immediate influence of the recreational drug on CD4 count whereas  the third and fourth    represent the distant influence. We see that the recreational drug has a decreasing  distant influence on CD4 count. The SAS code producing the results of Table $3c$ is presented in Supplementary Material B.

\section{Concluding Remarks}
In this article, we have shown that the point parameters -- i.e. point effects of treatments, point effects of covariates between consecutive treatments and a grand mean –-   form a point parametrization for the conditional distribution of an   outcome given  all treatments  and   time-dependent covariates.
In  point parametrization, we estimate sequential causal effects by maximum likelihood.
Using methods in  single-point causal inference (Rosenbaum  \&
Rubin 1983; Rosenbaum 1995; Rubin 2005), we improve
the   estimation    by    constraint on the point effects of treatments   and  assignment conditions of treatments.

Given data, an outcome model and the likelihood,  our estimation of sequential causal effects is most efficient due to the nature of maximum likelihood. The point estimation  is  unbiased for finite sample while the interval estimation    achieves the  nominal  coverage probability.
Furthermore, the ML estimates of sequential causal effects are consistent in many practical situations, where the underlying net effects   have a pattern of finite dimension while  treatment variables and covariates take finite numbers of values.
The consistency is true even when the treatment sequence  gets long and the number of point parameters increases exponentially.
It is   interesting to compare this consistency with  the inconsistency of the ML estimate of the causal effect of a single-point treatment in adjustment of a confounder of infinite dimension (Robins \& Ritov 1997). In the latter case, the ML estimate of the   causal  effect of a single-point treatment is highly correlated with that of the confounder of infinite dimension.

There are rich literatures about  semi-parametric approaches to sequential causal inference, for instance, the marginal structural model and the structural nested model. These  approaches can deal with more complex problems but use  likelihoods with additional  assumptions: the marginal structural model is based on a weighted likelihood while the structural nested model is based on the likelihood of treatment assignment. In comparison, our approach only uses a genuine likelihood, though in a relatively simple setting.

Due to the scope of this article, we  have only considered the following  setting:  treatments and   covariates are discrete,  the outcome model is linear,   the point and net effects of treatments and the sequential causal effects are measured by differences, and the variability of treatments and covariates is ignored.
On the other hand,
  methods   are  available to estimate the causal effect of  one single-point treatment in  complex settings. We believe that
analogous methods can be developed to estimate  sequential causal effects in  complex settings.

\bigskip
\begin{center}
{\large\bf SUPPLEMENTARY MATERIALS}
\end{center}
\begin{description}

\item[Supplementary material A:] Proofs for formulas  (\ref{na11}), (\ref{neq1_0}) and (\ref{n_1_0_0_1}) and Theorem \ref{nt1}

\item[Supplementary material  B:] (1) SAS codes  and SAS data sets   for the simulation study in Section $5.2$ and (2) SAS code   and  SAS data set for the illustration with a HIV study in Section $5.3$. (Zip file)

\end{description}

\pagebreak

\section*{Parametric Sequential Causal Inference in Point Parametrization: Supplementary Material A}
\begin{center}
Li Yin and Xiaoqin Wang
\end{center}

\subsection*{Proof of  formula  (\ref{na11}) in Section $3.1$ of the article}
Formula (\ref{nn1})  in Section $3.1$ of the article is written as
$$
\mu(\mathbf
z_1^{t-1}, \mathbf x_1^{t-1}, z_t )=\mu(\mathbf z_1^{t-1}, \mathbf
x_1^{t-1}, z_t=0)+\theta(\mathbf z_1^{t-1}, \mathbf x_1^{t-1}, z_t),
$$
where  we take $\theta(\mathbf z_1^{t-1},\mathbf x_1^{t-1},
z_t=0)=0$.  At $t=T$, this becomes
\begin{equation}\label{a2_1_1}
\mu(\mathbf z_1^{T},\mathbf x_1^{T-1})=\mu(\mathbf z_1^{T-1},\mathbf
x_1^{T-1}, z_{T}=0) + \theta(\mathbf z_1^{T-1},\mathbf x_1^{T-1},
z_T).
\end{equation}
Taking   average on both sides of (\ref{a2_1_1}) with respect to
${\rm pr}(z_T\mid \mathbf
z_1^{T-1},\mathbf x_1^{T-1})$, we obtain
$$\mu(\mathbf z_1^{T-1},\mathbf x_1^{T-1})=\mu(\mathbf
z_1^{T-1},\mathbf x_1^{T-1}, z_{T}=0) + \sum_{z_T^*}\theta(\mathbf
z_1^{T-1},\mathbf x_1^{T-1}, z_T^*) {\rm pr}(z_T^*\mid \mathbf
z_1^{T-1},\mathbf x_1^{T-1}),$$
which  implies
$$
\mu(\mathbf
z_1^{T-1},\mathbf x_1^{T-1}, z_{T}=0)=-\sum_{z_T^*}\theta(\mathbf
z_1^{T-1},\mathbf x_1^{T-1}, z_T^*){\rm pr}(z_T^*\mid \mathbf
z_1^{T-1},\mathbf x_1^{T-1})+\mu(\mathbf z_1^{T-1},\mathbf
x_1^{T-1}).
$$
 Inserting this     into (\ref{a2_1_1}), we obtain
\begin{equation}\label{a2_1_1_1}
\mu(\mathbf z_1^{T},\mathbf x_1^{T-1})=
\end{equation}
$$\sum_{z_T^* }
-\theta(\mathbf z_1^{T-1},\mathbf x_1^{T-1}, z_T^*){\rm
pr}(z_T^*\mid \mathbf z_1^{T-1},\mathbf x_1^{T-1}) +\theta(\mathbf
z_1^{T-1},\mathbf x_1^{T-1}, z_T) + \mu(\mathbf z_1^{T-1},\mathbf
x_1^{T-1}).
$$

Formula (\ref{nn2})  in Section $3.1$ of the article is written as
$$
\mu(\mathbf
z_1^{t}, \mathbf x_1^{t-1}, \mathbf x_{t}  )=\mu(\mathbf z_1^{t},
\mathbf x_1^{t-1}, \mathbf x_{t}=\mathbf 0)+\gamma(\mathbf z_1^{t}, \mathbf x_1^{t-1},\mathbf x_{t}),
$$
where  we take $\gamma(\mathbf z_1^{t},\mathbf x_1^{t-1},
\mathbf x_t=\mathbf 0)=  0$.
Using this formula at $t=T-1$ and then following the above procedure, we obtain
\begin{equation}\label{a2_1_1_2}
\mu(\mathbf z_1^{T-1},\mathbf x_1^{T-1})=
\end{equation}
$$\sum_{\mathbf x_{T-1}^* }
-\gamma(\mathbf z_1^{T-1},\mathbf x_1^{T-2}, \mathbf x_{T-1}^*){\rm
pr}(\mathbf x_{T-1}^*\mid \mathbf z_1^{T-1},\mathbf x_1^{T-2})
+\gamma(\mathbf z_1^{T-1},\mathbf x_1^{T-2}, \mathbf x_{T-1}) +
\mu(\mathbf z_1^{T-1},\mathbf x_1^{T-2}).
$$

Inserting (\ref{a2_1_1_2}) into (\ref{a2_1_1_1}), we obtain
$$
\mu(\mathbf z_1^{T},\mathbf x_1^{T-1})=\sum_{z_T^* }
-\theta(\mathbf z_1^{T-1},\mathbf x_1^{T-1}, z_T^*){\rm
pr}(z_T^*\mid \mathbf z_1^{T-1},\mathbf x_1^{T-1}) +\theta(\mathbf
z_1^{T-1},\mathbf x_1^{T-1}, z_T)
$$
$$
+\sum_{\mathbf x_{T-1}^* }
-\gamma(\mathbf z_1^{T-1},\mathbf x_1^{T-2}, \mathbf x_{T-1}^*){\rm
pr}(\mathbf x_{T-1}^*\mid \mathbf z_1^{T-1},\mathbf x_1^{T-2})
+\gamma(\mathbf z_1^{T-1},\mathbf x_1^{T-2}, \mathbf x_{T-1})
$$
$$
+\mu(\mathbf z_1^{T-1},\mathbf x_1^{T-2}).
$$
We go on with the same procedure for $\mu(\mathbf z_1^{T-1},\mathbf
x_1^{T-2}),\ldots,\mu(z_1)$ consecutively and  finally  obtain
$$
\mu(\mathbf z_1^T,\mathbf x_1^{T-1}) =\sum_{t=1}^T \left
[\sum_{z_t^*} - \theta(\mathbf z_1^{t-1},\mathbf x_1^{t-1},
z_t^*){\rm pr}(z_t^*\mid \mathbf z_1^{t-1},\mathbf x_1^{t-1}) +
\theta(\mathbf z_1^{t-1},\mathbf x_1^{t-1}, z_t)\right ]+
$$
$$
\sum_{t=1}^{T-1} \left [\sum_{\mathbf x_{t}^*}
-\gamma(\mathbf z_1^{t},\mathbf x_1^{t-1}, \mathbf x_{t}^*){\rm
pr}(\mathbf x_{t}^*\mid \mathbf z_1^{t},\mathbf x_1^{t-1})+
\gamma(\mathbf z_1^{t},\mathbf x_1^{t-1}, \mathbf x_{t})\right ]+
\mu,
$$
which is (\ref{na11})   of Section $3.1$ in the article.

Formula (\ref{a2_1_1_1}) is true for any $T$. Taking $T=t$ and replacing $z_t$ by $z_t^*$ and $z_t^*$ by $z_t^{**}$,  we obtain
\begin{equation}\label{a2_1_1_3_0}
\mu(\mathbf z_1^{t-1},\mathbf x_1^{t-1},z_t^*) =
\end{equation}
$$
 \sum_{z_t^{**}} -
\theta(\mathbf z_1^{t-1},\mathbf x_1^{t-1}, z_t^{**}){\rm pr}(z_t^{**}\mid
\mathbf z_1^{t-1},\mathbf x_1^{t-1}) + \theta(\mathbf
z_1^{t-1},\mathbf x_1^{t-1}, z_t^*)
+ \mu(\mathbf z_1^{t-1},\mathbf
x_1^{t-1}),
$$
which will be used below in the proof of Theorem  \ref{nt1} in Section $4.2$ of the article.

 \subsection*{Proof of formula (\ref{neq1_0}) in Section $3.2$ of the article}
 The first part of assumption (\ref{eq1}) in Section $2.2$ of the article is
 $$
  \mathbf x_t^{T-1}(\mathbf z_t^{T-1}),
y(\mathbf z_t^{T})  \bot z^{*}_t \mid \mathbf z_1^{t-1}, \mathbf
x_1^{t-1} ,
 $$
 which
at  $t=T$   is
$$
y(z_T)  \bot z^{*}_{T} \mid \mathbf z_1^{T-1}, \mathbf
x_1^{T-1},
$$
which implies
\begin{equation}\label{a3_1_0_0_0}
E\{y( z_T
)\mid \mathbf z_1^{T-1},\mathbf x_1^{T-1}\} =E\{y(z_T)\mid \mathbf z_1^{T-1},\mathbf x_1^{T-1}, z_T\}
\end{equation}
$$
=E\{y\mid \mathbf z_1^{T-1},\mathbf x_1^{T-1}, z_T\}= \mu(\mathbf z_1^{T},\mathbf x_1^{T-1} ).
$$
Definition
(\ref{n_1_0_0}) in Section $3.2$ of the article is written as
$$
E\{y( z_t, \mathbf z_{t+1}^{T}=\mathbf 0)\mid
\mathbf z_1^{t-1}, \mathbf x_1^{t-1}\}=E\{y( z_t =0, \mathbf
z_{t+1}^{T}=\mathbf 0)\mid \mathbf z_1^{t-1}, \mathbf x_1^{t-1}\}+\phi(\mathbf z_1^{t-1}, \mathbf
x_1^{t-1},z_t),
$$
which at $t=T$ is
$$
E\{y( z_T)\mid
\mathbf z_1^{T-1}, \mathbf x_1^{T-1}\}=E\{y( z_T =0)\mid \mathbf z_1^{T-1}, \mathbf x_1^{T-1}\}+\phi(\mathbf z_1^{T-1}, \mathbf
x_1^{T-1},z_T).
$$
Combining this  with
(\ref{a3_1_0_0_0}), we obtain
$$\mu(\mathbf z_1^{T},\mathbf x_1^{T-1} )=E\{y( z_T= 0
)\mid \mathbf z_1^{T-1},\mathbf x_1^{T-1}\}+\phi(  \mathbf
z_1^{T-1},\mathbf{x}_1^{T-1},z_T),
$$
which is  (\ref{neq1_0}) of the article for  $t=T$.

Now we derive (\ref{neq1_0}) for $t=1,\ldots, T-1$.
Formula (\ref{nn1_0}) in Section $3.1$ of the article is
$$
\mu(\mathbf z_1^{t}, \mathbf x_1^{t-1})= \sum_{\mathbf z_{t+1}^{T},
\mathbf
x_{t}^{T-1}} \mu( \mathbf z_1^{T},\mathbf
x_1^{T-1}) {\rm pr}(\mathbf z_{t+1}^{T},\mathbf x_{t}^{T-1}\mid
\mathbf z_1^{t},\mathbf x_1^{t-1}).
$$
Inserting (\ref{a3_1_0_0_0})  into  this, we
obtain
\begin{equation}\label{a3_1_0_0}
\mu(\mathbf z_1^{t}, \mathbf x_1^{t-1})= \sum_{\mathbf z_{t+1}^{T},
\mathbf
x_{t}^{T-1}}E\{y(z_T)\mid \mathbf
z_1^{T-1},\mathbf x_1^{T-1}\}{\rm pr}(\mathbf
z_{t+1}^{T},\mathbf x_{t}^{T-1}\mid \mathbf z_1^{t},\mathbf x_1^{t-1}).
\end{equation}
Let  $A(t)=E\{y(z_t,\mathbf z_{t+1}^T=\mathbf 0)\mid \mathbf
z_1^{t-1},\mathbf x_1^{t-1}\}$ and
$$
A(s)=\sum_{\mathbf z_{t+1}^{s},\mathbf x_{t}^{s-1}}
E\{y(z_s,\mathbf z_{s+1}^T=\mathbf 0)\mid \mathbf z_1^{s-1},\mathbf
x_1^{s-1}\}{\rm pr}(\mathbf z_{t+1}^{s},\mathbf x_{t}^{s-1}\mid
\mathbf z_1^{t},\mathbf x_1^{t-1})
$$
for $s=t+1,\ldots, T$.

Comparing (\ref{a3_1_0_0}) with $A(T)$, we see that $\mu(\mathbf z_1^{t}, \mathbf
x_1^{t-1})=A(T)$, which is written as
$$
A(T)=
$$
$$\sum_{\mathbf z_{t+1}^{T},\mathbf x_{t}^{T-1}}[E\{y(z_T)\mid
\mathbf z_1^{T-1},\mathbf x_1^{T-1}\}-E\{y(z_T=0)\mid \mathbf
z_1^{T-1},\mathbf x_1^{T-1}\}]{\rm pr}(\mathbf
z_{t+1}^{T},\mathbf x_{t}^{T-1}\mid \mathbf z_1^{t},\mathbf x_1^{t-1})
$$
\begin{equation}\label{a3_1_0}
+\sum_{\mathbf z_{t+1}^{T},\mathbf x_{t}^{T-1}}E\{y(z_T=0)\mid
\mathbf z_1^{T-1},\mathbf x_1^{T-1}\}{\rm pr}(\mathbf z_{t+1}^{T},
\mathbf
x_{t}^{T-1}\mid \mathbf z_1^{t},\mathbf
x_1^{t-1})
\end{equation}
$$
= \sum_{\mathbf
z_{t+1}^{T-1},\mathbf x_{t}^{T-1}}\sum_{z_T>0}\phi(\mathbf z_1^{T-1}, \mathbf
x_1^{T-1},z_T){\rm pr}(\mathbf z_{t+1}^{T-1},\mathbf x_{t}^{T-1},z_T\mid
\mathbf z_1^{t},\mathbf x_1^{t-1})
$$
\begin{equation}\label{a3_1_0_1}
+\sum_{\mathbf z_{t+1}^{T-1},\mathbf x_{t}^{T-2}}E\{y(z_T=0)\mid
\mathbf z_1^{T-1},\mathbf x_1^{T-2}\}{\rm pr}(\mathbf z_{t+1}^{T-1},
\mathbf
x_{t}^{T-2}\mid \mathbf z_1^{t},\mathbf
x_1^{t-1}).
\end{equation}
Here  the first summation term in (\ref{a3_1_0})  is equal to the first summation term in  (\ref{a3_1_0_1}) according to definition
(\ref{n_1_0_0}) at  $t=T$; the second summation term in  (\ref{a3_1_0}), after being summed up over $z_T$ and $\mathbf x_{T-1}$, is equal  to the second summation term in (\ref{a3_1_0_1}).

The first part of assumption (\ref{eq1}) of the article for  $t=T-1$  is
$$
y(z_{T-1},z_T)  \bot z^{*}_{T-1} \mid \mathbf z_1^{T-2}, \mathbf
x_1^{T-2},
$$
which implies
$$
E\{y(z_{T-1}, z_T=0)\mid \mathbf
z_1^{T-2},\mathbf x_1^{T-2}\}=E\{y(z_{T-1}, z_T=0)\mid \mathbf
z_1^{T-2},\mathbf x_1^{T-2},z_{T-1}\}
$$
\begin{equation}\label{a3_1_1_0}
= E\{y(z_T=0)\mid
\mathbf z_1^{T-2},\mathbf x_1^{T-2}, z_{T-1}\}.
\end{equation}
Thus
the
second summation term in (\ref{a3_1_0_1}) is equal to
$$\sum_{\mathbf z_{t+1}^{T-1}
,\mathbf
x_{t}^{T-2}}E\{y(z_{T-1}, z_T=0)\mid \mathbf
z_1^{T-2},\mathbf x_1^{T-2}\}{\rm pr}(\mathbf
z_{t+1}^{T-1},\mathbf x_{t}^{T-2}\mid \mathbf z_1^{t},\mathbf x_1^{t-1}),
$$
 which is
$A(T-1)$.

Hence we obtain
\begin{equation}\label{a3_1_1}
A(T)=\sum_{\mathbf z_{t+1}^{T-1},\mathbf x_{t}^{T-1}} \sum_{z_T>0}
\phi(\mathbf z_1^{T-1}, \mathbf x_1^{T-1},z_T) {\rm pr}(\mathbf z_{t+1}^{T-1},
\mathbf
x_{t}^{T-1},z_T\mid \mathbf z_1^{t},\mathbf
x_1^{t-1})+A(T-1).
\end{equation}
We continue with  the same procedure to rewrite  $A(T-1)$,$\ldots$,
$A(t+1)$  consecutively and then
$$
A(t)= E\{y(z_t,\mathbf z_{t+1}^T=\mathbf 0)\mid \mathbf
z_1^{t-1},\mathbf x_1^{t-1}\}-E\{y(z_t=0,\mathbf z_{t+1}^T=\mathbf 0)\mid \mathbf
z_1^{t-1},\mathbf x_1^{t-1}\}
$$
$$
+E\{y(z_t=0,\mathbf z_{t+1}^T=\mathbf 0)\mid \mathbf
z_1^{t-1},\mathbf x_1^{t-1}\}=\phi(\mathbf z_1^{t-1}, \mathbf x_1^{t-1},z_t)+
E\{y(\mathbf z_{t}^T=\mathbf 0)\mid \mathbf
z_1^{t-1},\mathbf x_1^{t-1}\}.
$$
Finally we obtain
$$
A(T)=\sum_{s=t+1}^T\sum_{
\mathbf z_{t+1}^{s-1},\mathbf x_{t}^{s-1}}\sum_{z_s>0  }\phi(\mathbf z_1^{s-1}, \mathbf
x_1^{s-1},z_s){\rm pr}(
\mathbf z_{t+1}^{s-1},\mathbf x_{t}^{s-1},z_s\mid \mathbf
z_1^{t},\mathbf x_1^{t-1})
$$
$$
+\ \phi(\mathbf z_1^{t-1},\mathbf x_1^{t-1},z_t )+E\{y(\mathbf z_t^T=\mathbf 0
)\mid \mathbf z_1^{t-1},\mathbf x_1^{t-1}\},
$$
which is  (\ref{neq1_0}) of the article for $t=1,\ldots, T-1$ because $A(T)=\mu(\mathbf z_1^{t},\mathbf x_1^{t-1} )$.

\subsection*{Proof of  formula  (\ref{n_1_0_0_1}) in Section $3.3$ of the article}
The $G$-formula (\ref{eq8}) in Section $2.2$ of the article  is written as
$$
E\{y(\mathbf z_{1}^T)\}=\sum_{\mathbf x_1^{T-1}} \mu(\mathbf z_1^{T},\mathbf x_1^{T-1})
\prod_{s=1}^{T-1} {\rm pr}(\mathbf x_s\mid \mathbf z_1^{s},\mathbf
x_1^{s-1}).
$$
As shown by (\ref{a3_1_0_0_0}), assumption (\ref{eq1}) of the article implies
$$
\mu(\mathbf z_1^{T},\mathbf x_1^{T-1} ) = E\{y( z_T
)\mid \mathbf z_1^{T-1},\mathbf x_1^{T-1}\}.
$$
Inserting this into $E\{y(\mathbf z_{1}^T)\}$, we obtain
$$
E\{y(\mathbf z_{1}^T)\}=\sum_{\mathbf x_1^{T-1}} E\{y( z_T
)\mid \mathbf z_1^{T-1},\mathbf x_1^{T-1}\} \prod_{s=1}^{T-1}{\rm pr}(\mathbf x_s\mid \mathbf z_1^{s},\mathbf
x_1^{s-1}).
$$
Let  $C(1)= E\{y(z_1, \mathbf  z_2^T=\mathbf 0
)\}$ and
$$
C(t)=\sum_{\mathbf x_1^{t-1}} E\{y( z_t, \mathbf z_{t+1}^T=\mathbf 0
)\mid \mathbf z_1^{t-1},\mathbf x_1^{t-1}\} \prod_{s=1}^{t-1}{\rm pr}(\mathbf x_s\mid \mathbf z_1^{s},\mathbf
x_1^{s-1})
$$
for $t=2,\ldots, T$.
Then $C(T)=
E\{y(\mathbf z_{1}^T)\}$.

Using definition
(\ref{n_1_0_0}) of the article  at  $t=T$, we rewrite $C(T)$ as
$$
C(T)=\sum_{\mathbf x_1^{T-1}}\phi(\mathbf z_1^{T-1}, \mathbf
x_1^{T-1},z_T)\prod_{s=1}^{T-1} {\rm pr}(\mathbf x_s\mid \mathbf z_1^{s},\mathbf
x_1^{s-1})
$$
$$
+\sum_{\mathbf x_1^{T-1}}E\{y( z_T=  0
)\mid \mathbf z_1^{T-1},\mathbf x_1^{T-1}\}\prod_{s=1}^{T-1} {\rm pr}(\mathbf x_s\mid \mathbf z_1^{s},\mathbf
x_1^{s-1}).
$$
The last summation term, summing over $\mathbf x_{T-1}$ with respect to ${\rm pr}(\mathbf x_{T-1}\mid \mathbf z_1^{T-1},\mathbf
x_1^{T-2})$,  is equal to
$$
\sum_{\mathbf x_1^{T-2}}E\{y( z_T=  0
)\mid \mathbf z_1^{T-1},\mathbf x_1^{T-2}\}\prod_{s=1}^{T-2} {\rm pr}(\mathbf x_s\mid \mathbf z_1^{s},\mathbf
x_1^{s-1}).
$$
As shown by (\ref{a3_1_1_0}), assumption (\ref{eq1}) of the article  implies
$$
E\{y(z_T=0)\mid
\mathbf z_1^{T-2},\mathbf x_1^{T-2}, z_{T-1}\}=E\{y(z_{T-1}, z_T=0)\mid \mathbf
z_1^{T-2},\mathbf x_1^{T-2}\}.
$$
Therefore the above summation term  is equal to
$$
\sum_{\mathbf x_1^{T-2}}E\{y(z_{T-1}, z_T=0)\mid \mathbf
z_1^{T-2},\mathbf x_1^{T-2}\}\prod_{s=1}^{T-2} {\rm pr}(\mathbf x_s\mid \mathbf z_1^{s},\mathbf
x_1^{s-1}).
$$
which is equal to $C(T-1)$.

Hence we have
$$
C(T)=\sum_{\mathbf x_1^{T-1}}\phi(\mathbf z_1^{T-1}, \mathbf
x_1^{T-1},z_T)\prod_{s=1}^{T-1} {\rm pr}(\mathbf x_s\mid \mathbf z_1^{s},\mathbf
x_1^{s-1}) + C(T-1).
$$
We continue  with the same procedure to rewrite $C(T-1),\ldots, C(2)$ and then
$$
C(1)=\phi(z_1)+E\{y(\mathbf z_{1}^T=\mathbf 0)\} .
$$
Finally we
obtain
$$
E\{y(\mathbf z_{1}^T)\}=\sum_{t=2}^T\sum_{\mathbf x_1^{t-1}} \phi(\mathbf z_1^{t-1}, \mathbf
x_1^{t-1},z_t)
\prod_{s=1}^{t-1} {\rm pr}(\mathbf x_s\mid \mathbf z_1^{s},\mathbf
x_1^{s-1})  + \phi(z_1) +E\{y(\mathbf z_{1}^T=\mathbf 0)\},
$$
which is (\ref{n_1_0_0_1}) of the article.

\subsection*{Proof of Theorem  \ref{nt1} in Section $4.2$ of the article }
Formula (\ref{na11}) in Section $3.1$ of the article, proved earlier in this supplementary material, is written as
$$
\mu(\mathbf z_1^{*T},\mathbf x_1^{*(T-1)}) =
$$
$$
\sum_{t=1}^T \left
[\sum_{z_t^{**}} - \theta(\mathbf z_1^{*(t-1)},\mathbf x_1^{*(t-1)},
z_t^{**}){\rm pr}(z_t^{**}\mid \mathbf z_1^{*(t-1)},\mathbf x_1^{*(t-1)}) +
\theta(\mathbf z_1^{*(t-1)},\mathbf x_1^{*(t-1)}, z_t^*)\right ]
$$
$$
+\sum_{t=1}^{T-1} \left [\sum_{\mathbf x_{t}^{**}}
-\gamma(\mathbf z_1^{*t},\mathbf x_1^{*(t-1)}, \mathbf x_{t}^{**}){\rm
pr}(\mathbf x_{t}^{**}\mid \mathbf z_1^{*t},\mathbf x_1^{*(t-1)})+
\gamma(\mathbf z_1^{*t},\mathbf x_1^{*(t-1)}, \mathbf x_{t}^*)\right ]+
\mu.
$$
Its partial derivative with respect to $\theta(\mathbf z_1^{t-1},\mathbf
x_1^{t-1},z_t)$ is
$$
\frac{\partial
\mu(\mathbf z_1^{*T},\mathbf x_1^{*(T-1)}) }{\partial \theta(\mathbf z_1^{t-1},\mathbf
x_1^{t-1},z_t)}
=I_{(\mathbf z_1^{t-1},\mathbf x_1^{t-1})}(\mathbf
z_1^{*(t-1)},\mathbf x_1^{*(t-1)}) \{I_{z_t}(z_t^*)- {\rm
pr}(z_t\mid \mathbf z_1^{t-1},\mathbf x_1^{t-1}) \},
$$
 where
$I_a(b)$ takes  one if $b=a$ and   zero otherwise.
Let  $s(A)$ be the set of units in  stratum $A$ and $n(A)$ be the
number of units in   stratum $A$.
The score function   for the standard parameter $\mu(\mathbf z_1^{*T},\mathbf
x_1^{*(T-1)})$  is
$$
 U_{\mu(\mathbf z_1^{*T},\mathbf
x_1^{*(T-1)})}=\sum_{i\in s(\mathbf z_1^{*T},\mathbf
x_1^{*(T-1)})}\{y_i- \mu(\mathbf z_1^{*T},\mathbf x_1^{*(T-1)})\}.
$$

Using the Chain rule,
the score function  for the point parameter
$\theta(\mathbf
z_1^{t-1},\mathbf x_1^{t-1},z_t)$   is then
$$
U_{\theta(\mathbf z_1^{t-1},\mathbf x_1^{t-1},z_t)}=\sum_{\mathbf
z_1^{*T},\mathbf x_1^{*(T-1)}}U_{\mu(\mathbf z_1^{*T},\mathbf
x_1^{*(T-1)}) }\frac{\partial \mu(\mathbf z_1^{*T},\mathbf
x_1^{*(T-1)}) }{
\partial \theta(\mathbf z_1^{t-1},\mathbf
x_1^{t-1},z_t)}
$$
$$
=\sum_{\mathbf
z_1^{*T},\mathbf x_1^{*(T-1)}}
I_{(\mathbf z_1^{t-1},\mathbf x_1^{t-1})}(\mathbf
z_1^{*(t-1)},\mathbf x_1^{*(t-1)}) \{I_{z_t}(z_t^*)- {\rm
pr}(z_t\mid \mathbf z_1^{t-1},\mathbf x_1^{t-1}) \}
$$
$$
\sum_{i\in s(\mathbf z_1^{*T},\mathbf
x_1^{*(T-1)})}\{y_i- \mu(\mathbf z_1^{*T},\mathbf x_1^{*(T-1)})\}
$$
$$=\sum_{z_t^*}
\left \{I_{z_t}(z_t^*)-{\rm pr}(z_t\mid \mathbf z_1^{t-1},\mathbf
x_1^{t-1})\right \}\sum_{i\in s(\mathbf z_1^{t-1},\mathbf x_1^{t-1},
z_t^*)}\{ y_i-\mu(\mathbf z_1^{t-1},\mathbf x_1^{t-1}, z_t^*) \}
$$
$$
=\sum_{z_t^*}
\left \{I_{z_t}(z_t^*)-{\rm pr}(z_t\mid \mathbf z_1^{t-1},\mathbf
x_1^{t-1})\right \}\left\{
\sum_{i\in s(\mathbf z_1^{t-1},\mathbf x_1^{t-1},
z_t^*)}y_i -n(\mathbf z_1^{t-1},\mathbf x_1^{t-1}, z_t^*)
\mu(\mathbf z_1^{t-1},\mathbf x_1^{t-1}, z_t^*)
\right \}.
$$

Replacing $\mu(\mathbf z_1^{t-1},\mathbf x_1^{t-1}, z_t^*)$ by   (\ref{a2_1_1_3_0}) proved earlier in this supplementary material, we obtain
$$
U_{\theta(\mathbf z_1^{t-1},\mathbf x_1^{t-1}, z_t)}= \sum_{z_t^*}
\left \{I_{z_t}(z_t^*)-{\rm pr}(z_t\mid \mathbf z_1^{t-1},\mathbf
x_1^{t-1})\right \}  \left  [ \sum_{i\in s(\mathbf
z_1^{t-1},\mathbf x_1^{t-1}, z_t^*)} y_i  -n(\mathbf
z_1^{t-1},\mathbf x_1^{t-1},z_t^*)  \right .
$$
$$
\left \{ \sum_{z_t^{**}} -\theta(\mathbf z_1^{t-1},\mathbf
x_1^{t-1}, z_t^{**}){\rm pr}(z_t^{**}\mid \mathbf z_1^{t-1},\mathbf
x_1^{t-1}) + \theta(\mathbf z_1^{t-1},\mathbf x_1^{t-1},
z_t^*)\right \}
$$
$$
\left .  -n(\mathbf z_1^{t-1},\mathbf x_1^{t-1},z_t^*)\mu(\mathbf z_1^{t-1},\mathbf x_1^{t-1})
\right ].
$$

But we have
$$
\sum_{z_t^*}\left \{I_{z_t}(z_t^*)-{\rm pr}(z_t\mid \mathbf
z_1^{t-1},\mathbf x_1^{t-1})\right \} n(\mathbf z_1^{t-1},\mathbf
x_1^{t-1},z_t^*) \mu(\mathbf z_1^{t-1},\mathbf x_1^{t-1})=
$$
$$
\{n(\mathbf z_1^{t-1},\mathbf
x_1^{t-1},z_t)-{\rm pr}(z_t\mid \mathbf z_1^{t-1},\mathbf
x_1^{t-1})n(\mathbf z_1^{t-1},\mathbf
x_1^{t-1})\}\mu(\mathbf z_1^{t-1},\mathbf x_1^{t-1})=0.
$$
Therefore we  obtain
$$
U_{\theta(\mathbf z_1^{t-1},\mathbf x_1^{t-1}, z_t)}= \sum_{z_t^*}
\left \{I_{z_t}(z_t^*)-{\rm pr}(z_t\mid \mathbf z_1^{t-1},\mathbf
x_1^{t-1})\right \}  \left [ \sum_{i\in s(\mathbf
z_1^{t-1},\mathbf x_1^{t-1}, z_t^*)} y_i  -n(\mathbf
z_1^{t-1},\mathbf x_1^{t-1},z_t^*) \right .
$$
\begin{equation}\label{na1}
\left . \left \{ \sum_{z_t^{**}} -\theta(\mathbf z_1^{t-1},\mathbf
x_1^{t-1}, z_t^{**}){\rm pr}(z_t^{**}\mid \mathbf z_1^{t-1},\mathbf
x_1^{t-1}) + \theta(\mathbf z_1^{t-1},\mathbf x_1^{t-1},
z_t^*)\right \}\right ].
\end{equation}
From this formula, we see that $U_{\theta(\mathbf z_1^{t-1},\mathbf x_1^{t-1},z_t)}$
depends only on $\theta(\mathbf z_1^{t-1},\mathbf x_1^{t-1},z_t^*)$, thus proving Theorem \ref{nt1} of the article.

\end{document}